\PassOptionsToPackage{breaklinks=true}{hyperref}
\documentclass[fleqn,usenatbib]{mnras}

\usepackage{newtxtext,newtxmath}

\usepackage[T1]{fontenc}
\usepackage{ae,aecompl}

\usepackage{graphicx}	
\usepackage{amsmath}	
\usepackage{amssymb}	
\usepackage{float}
\usepackage[caption = false]{subfig}

\newcommand*{\transpose}{^{\mkern-1.5mu\mathsf{T}}}

\title[Automatic classification of tidal debris]{A machine--vision method for automatic classification of stellar halo substructure}

\author[D. Hendel et al.]{
David Hendel$^{1,2}$\thanks{E-mail: hendel@astro.columbia.edu}, Kathryn V. Johnston$^{1}$, Rohit K. Patra$^{3}$, and Bodhisattva Sen$^{4}$
\\
$^{1}$Department of Astronomy, Columbia University, 550 West 120th Street, New York, NY 10027, USA\\
$^{2}$Department of Astronomy and Astrophysics, University of Toronto, 50 St. George Street, Toronto, ON, M5S 3H4, Canada\\
$^{3}$Department of Statistics, University of Florida, 221 Griffin-Floyd Hall, Gainesville, FL 32611, USA\\
$^{4}$Department of Statistics, Columbia University, 1255 Amsterdam Avenue, New York, NY 10027, USA\\
}

\date{Accepted XXX. Received YYY; in original form ZZZ}

\pubyear{2018}

\begin{document}
\label{firstpage}
\pagerange{\pageref{firstpage}--\pageref{lastpage}}
\maketitle

\begin{abstract}
Tidal debris structures formed from disrupted satellites contain important clues about the assembly histories of galaxies. To date, studies of these structures have been hampered by reliance on by--eye identification and morphological classification which leaves their interpretation significantly uncertain. In this work we present a new machine-vision technique based on the Subspace--Constrained Mean Shift (SCMS) algorithm which can perform these tasks automatically. SCMS finds the location of the high-density `ridges' that define substructure morphology. After identification, the coefficients of an orthogonal series density estimator are used to classify points on the ridges as part of a continuum between shell--like or stream--like debris, from which a global morphological classification can be determined. We dub this procedure Subspace Constrained Unsupervised Detection of Structure (SCUDS). By applying this tool to controlled N--body simulations of minor mergers we demonstrate that the extracted classifications correspond to the well--understood underlying physics of phase mixing. The application of SCUDS to resolved stellar population data from near--future surveys will inform our understanding of the buildup of galaxies stellar halos.
\end{abstract}





\section{Introduction}

In the prevailing cosmological picture structure at all scales forms due the gravitational collapse of overdensities left over after inflation. After an initial collapse into gravitationally bound halos, structure formation proceeds hierarchically in that large systems are built over time from assemblages of smaller systems as they collide and merge together \citep[e.g.~][]{1978MNRAS.183..341W}. This process continues in the modern epoch as evidenced by observations of satellite galaxies being tidally disrupted and eventually subsumed into their hosts. The morphology of the stellar debris left over from these events can persist for many billions of years before phase--mixing into the smooth stellar halo \citep[e.g.~][]{1999MNRAS.307..495H} and the detailed structure of the unmixed debris contains a wealth of information about galactic merger histories \citep[c.f.][]{2008ApJ...689..936J}.

Two classic examples of ongoing mergers in the Local Group are the Sagittarius dwarf and its tidal tails around the Milky Way \citep{1994Natur.370..194I, 2002ApJ...569..245N, 2003ApJ...599.1082M, 2006ApJ...642L.137B} and the Giant Southern Stream in Andromeda \citep{2001Natur.412...49I,2002AJ....124.1452F}. These examples vividly demonstrate the ongoing nature of galactic accretion. In addition, a growing number of extragalactic surveys that reach very low surface brightness limits in integrated light show that similar structures are common around galaxies more generally \citep[e.g.~][]{2010AJ....140..962M, 2013ApJ...765...28A, 2015MNRAS.446..120D, 2017arXiv170904474G, 2018ApJ...857..144H, 2018arXiv180403330M, 2018arXiv180505970K} even though these studies are likely accessing only the bright end of the tidal debris luminosity function \citep{2005ApJ...635..931B, 2010MNRAS.406..744C}. Currently, the best samples contain a few hundred to a few thousand examples. In the near future these will be dwarfed by the catalogs produced by the full Hyper Suprime--Cam Subaru Strategic Program Wide Layer \citep[HSC-SSP,][]{2018PASJ...70S...4A}, which can detect low surface brightness galaxies and structures down to $\sim 27$ mag arcsec$^{-2}$  \citep{2018ApJ...857..104G, 2018arXiv180505970K}, and on a longer time baseline that of the upcoming Large Synoptic Survey Telescope (LSST) which is expected to reach a similar or slightly greater depth over more than ten times the sky area \citep{2009arXiv0912.0201L}. 

In addition there is a thriving literature on resolved stellar halos which also reveal extensive substructure. These studies -- which directly image individual stars at up to Mpc distances -- are very challenging but permit exceedingly low effective surface brightness limits and better determination of stellar ages and metallicities if they include multiband photometry. Pioneered by the PAndAS survey of Andromeda \citep{2007ApJ...671.1591I, 2009Natur.461...66M, 2014ApJ...780..128I}, a combination of ground--based, wide--field imaging studies covering tens to hundreds of projected kpc of nearby galaxies (\citealt{2010ApJ...714L..12M, 2011ApJ...736...24B, 2012MNRAS.419.1489B, 2014A&A...562A..73G, 2015ApJ...809L...1O}; 
PISCeS, \citealt{2016ApJ...823...19C}) and deep {\em Hubble Space Telescope} observations of select fields (\citealt{2009ApJS..183...67D}; GHOSTS, \citealt{2011ApJS..195...18R, 2014ApJ...791L...2R, 2016MNRAS.457.1419M, 2018arXiv180606828M}) are rapidly advancing our understanding of stellar halos and the substructure within them. 

LSST will also contribute to this work on resolved stellar halos, mapping individual red giant branch stars in the halos of galaxies out to approximately 6 Mpc after ten years of imaging. For approximately ten nearby large galaxies the imaging will be deep enough to obtain photometric metallicities as well, enabling more detailed studies of the stars' origins \citep{2009arXiv0912.0201L}. The next great leap forward is likely to come with the launch of the {\em Wide--Field Infrared Survey Telescope} ({\em WFIRST}). Combining a 0.34 deg$^{2}$ field of view, {\em HST} resolution, and an effective surface brightness limit of 35 mag arcsec$^{-2}$, {\em WFIRST} will be able to map stellar halos in resolved stellar populations with unprecedented detail. The {\em WFIRST} Infrared Nearby Galaxy Survey Science Investigation Team is planning to image most ($\sim 100$) large galaxies within 10 Mpc, resolving $\sim$ billions of stars including an expected tens of millions in their halos. This will be an ideal dataset to understand the underlying merger dynamics but requires substantial theoretical development. 

One way to extract astrophysical information from catalogs of substructure is to classify systems by their spatial morphology. Tidal debris can be broadly divided into two morphological groups: streams and shells. Streams are long, narrow features that approximate orbits in the host galaxy's potential. Shells, on the other hand, often cover a large two--dimensional area with significant luminosity, bounded at some radius from the host galaxy by a bright caustic that corresponds roughly to an isopotential. To first order streams and shells are generated by near--circular and near--radial satellite orbits, respectively, but in detail contain information about all the interaction parameters including the host mass, merger mass ratio, and interaction duration \citep{1984ApJ...279..596Q, 2008ApJ...689..936J, 2015MNRAS.450..575A, 2015MNRAS.454.2472H, 2018arXiv180810454K}. However, the catalogs that exist rely on visual identification (with the notable exception of \citealt{2018arXiv180505970K}) followed by manual classification into morphological groups. This introduces a host of worrying systematics that can hamper physical interpretation. Besides improving the uniformity and reproducibility of the resulting catalogs, algorithmic methods can also identify subtle structures and groupings that may not be obvious to observers even in extremely well--studied fields \citep[such as the M31 halo,][]{2018arXiv181008234M}.

In this contribution we seek to improve this state by introducing an automatic method for classifying tidal debris based on the Subspace--Constrained Mean Shift algorithm. In Section~\ref{sec:methods}, we describe the test dataset of N-body simulations and the classification pipeline; in Section~\ref{sec:results} we examine its performance compared to a semi--analytical `morphology metric' that we take to represent the true state of the interactions; and Section~\ref{sec:discussion} considers the possibilities for future use. Section~\ref{sec:conclusions} concludes.

\begin{figure}
	\includegraphics[width=\columnwidth]{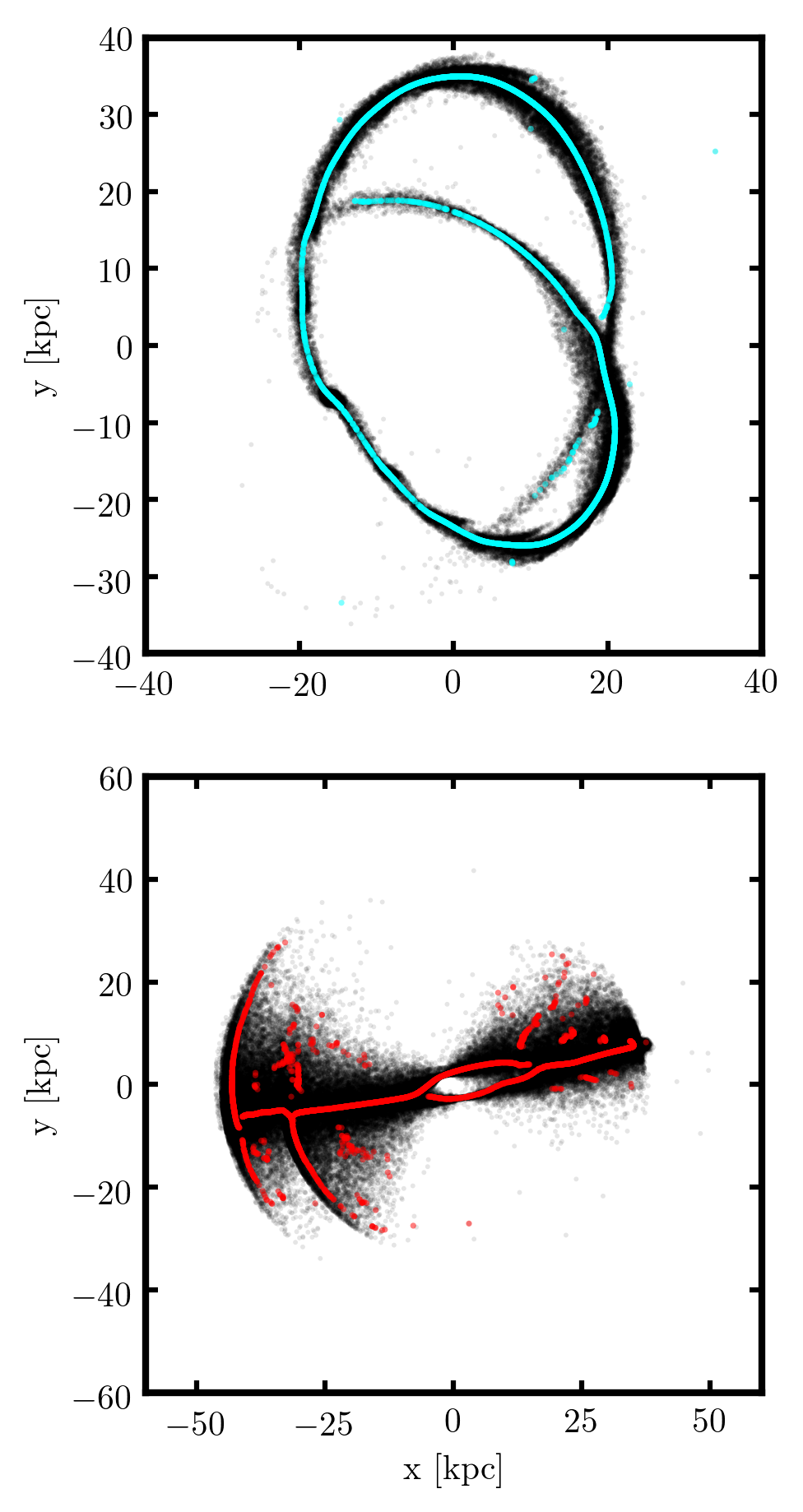}
        \caption{Ridge points computed with the Subspace--Constrained Mean Shift algorithm using a bandwidth of 2 kpc (colored points) plotted on top of the particle data (black points) for stream--forming (top) and shell--forming (bottom) minor merger simulations.  Both simulations shown are on orbits of the same energy as a circular orbit at 25 kpc, have a satellite mass of $6.5 \times 10^7\ {\rm{M_\odot}}$, and are integrated for 4.8 Gyr, but they have differing circularities of 0.90 (top) and 0.20 (bottom). Both the tidal streams and shell caustics are captured by the principal curves.}
    \label{fig:scms_example}
\end{figure}

\section{Algorithm: SCUDS}
\label{sec:methods}
Due to the advantages, rapidly advancing state, and future prospects of resolved maps of stellar halos, we have implemented an algorithm that operates on particle data. 
In this Section we describe the test data and pipeline, which can be separated into several stages: feature identification, feature classification, spatial reconstruction and finally debris classification. Each is described in turn below. Central to this method is the Subspace--Constrained Mean Shift algorithm \citep{SCMS}, so we dub our algorithm Subspace Constrained Unsupervised Detection of Structure, or SCUDS\footnote{One definition of {\em scud} is `loose vapory clouds driven swiftly by the wind,' which we find an apt visual.} for short.

\subsection{Test dataset}

N-body simulations provide an ideal proving ground for a machine--vision classifier since one can attempt to recover the precisely--known initial conditions. While in principle the output of any number of modern, public simulation datasets could be used, substantial particle resolution in the satellite galaxy is required, a wide variety of interaction parameters should be investigated, and it is convenient to have ready--made classifications to compare against. The grid of minor merger simulations 
used in \cite{2015MNRAS.454.2472H} satisfies these conditions.

The simulation setup evolves $10^5$--particle, NFW--profile \citep{1997ApJ...490..493N} satellites orbiting in a static, spherically symmetric NFW--profile host galaxy using the Self--Consistent Field basis function expansion code \citep{1992ApJ...386..375H}. The host potential has a viral mass of $1.77 \times 10^{12}\ \mathrm{M_\odot}$ and a scale radius of 24.6 kpc, broadly consistent with the Milky Way, but besides a small effect from the mass-concentration relation the mergers are nearly scale invariant with respect to $\mathrm{M_{host}}$. In total, this archive holds 1,920 snapshots representing mergers over a wide range of orbits (with energies equal to those of a circular orbit at 25, 45, 75, and 100 kpc, and 12 circularites between 0.05 and 0.95), mass ratios (NFW satellites of masses m/M$_\odot = 6.5 \times 10^6,10^7,10^8,10^9)$ and interaction times (up to 8 Gyr), as well as a morphological classification derived from semi-analytic scalings that describe the tidal disruption process (their Eq 5--8 and Figure 7). 

Even though these snapshots have a large satellite particle count compared to mergers with a similar mass ratio in modern cosmological hydrodynamic simulations, some initial conditions still produce debris structures with very low densities which makes any analysis uncertain at best. For this reason we will focus on the more tightly bound and lower mass examples, which maintain reasonable projected particle densities for many Gyr. Applying SCUDS to fully observationally--motivated merger ratios and orbit distributions simply requires additional resolution, which we defer to future work. Similarly, instead of tagging the most bound N-body particles as those representing stars, we assume all particles will be visible to retain sufficient data on which to operate. This means that the satellites do not exactly match the observed mass--concentration relation but again this is unimportant in terms of analyzing the debris structures; one can think of the simulation as representing a satellite whose dark halo has already been substantially stripped, which is a likely prerequisite for forming any kind of tidal debris \citep{2008ApJ...673..226P, 2012MNRAS.424.2401V, 2013MNRAS.431.3533C, 2016ApJ...833..109S}.

\begin{figure*}
	\includegraphics[width=\linewidth]{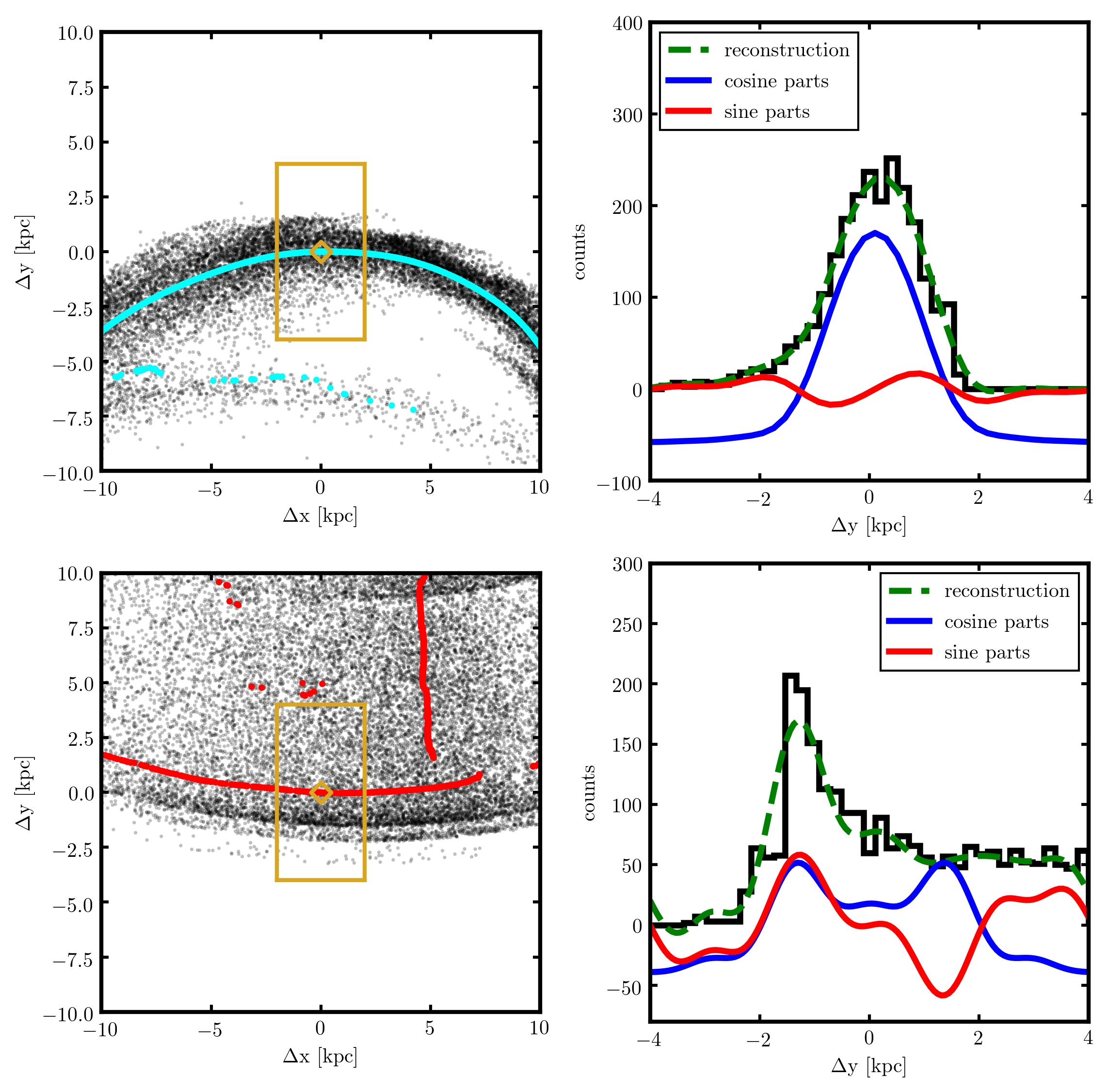}
        \caption{Methodology for classifying individual ridge points.  An example of a point on a stream (top) and shell (bottom) are shown. Left: A single $\vec{x}_r$ (gold diamond) is chosen and the local ridge is axis--aligned by a rotation such that the corresponding $\vec{\phi}_r$ points along the x--axis. A selection box (gold rectangle) is constructed for the particle data using these coordinates. Right: the data in the selection box is collapsed in the $x$ coordinate and then used to evaluate the coefficients in an orthogonal basis density estimator (Equation~\ref{eq:coeff}). The resulting density estimate is shown (Equation~\ref{eq:estimator}, green dashed line), appropriately scaled to match the density histogram, as well as the contributions from the symmetric cosine (blue line) and asymmetric sine (red line) components.}
    \label{fig:pointclassifier}
\end{figure*}

\subsection{Machine vision method}
\subsubsection{Feature identification: Subspace--Constrained Mean Shift}

The Mean Shift technique \citep{meanshift} is well--known as a non--parametric method of classification. Given a dataset $\{ {\vec{X}_i} \}$, an estimate of its density field $\rho(\vec{x})$ in $D$ dimensions, and a test point initial condition $\vec{y}$, Mean Shift iteratively moves $\vec{y}$ up the local density gradient towards a local maximum until convergence at a mode, where updates give shifts less than a chosen tolerance $\epsilon$. In the common use case as a classifier this is performed for each $\vec{y} \in \{ {\vec{X}_i} \}$ and all points that reach the same mode (within $\epsilon$) are considered a single cluster. However, for detection of tidal features we are interested not in the countable set of modes but rather the {\em principal curves}, by which we mean smooth curves (or manifolds) that pass locally through the middle of the data \citep{HastieThesis, Hastie1989}. These curves will trace the high-density regions of tidal streams and shell caustics, and we will refer to them as {\em density ridges} or {\em ridgelines}. Mean Shift can be extended with an additional constraint on the direction that $\vec{y}$ is allowed to move at each iteration (within a local subspace, thus Subspace--Constrained Mean Shift, or SCMS) so that it converges to just such a principal surface, of dimension $\ d \leq D-1$ \citep{SCMS}. This technique has been used successfully to identify filaments in the cosmic web, which have a similar geometry \citep{2015MNRAS.454.1140C, 2015MNRAS.454.3341C, 2016MNRAS.461.3896C, 2017MNRAS.466.1880C, 2017arXiv170902543H}. The specific implementation of SCMS we employ is distributed in the helit\footnote{https://github.com/thaines/helit/} package. In the following we assume $D=2$, i.e.~using sky positions alone. This could be extended to include a third dimension such as metallicity to separate stars that are part of tidal debris from the stellar halo of the host galaxy. 

\begin{figure*}
	\includegraphics[width=\linewidth]{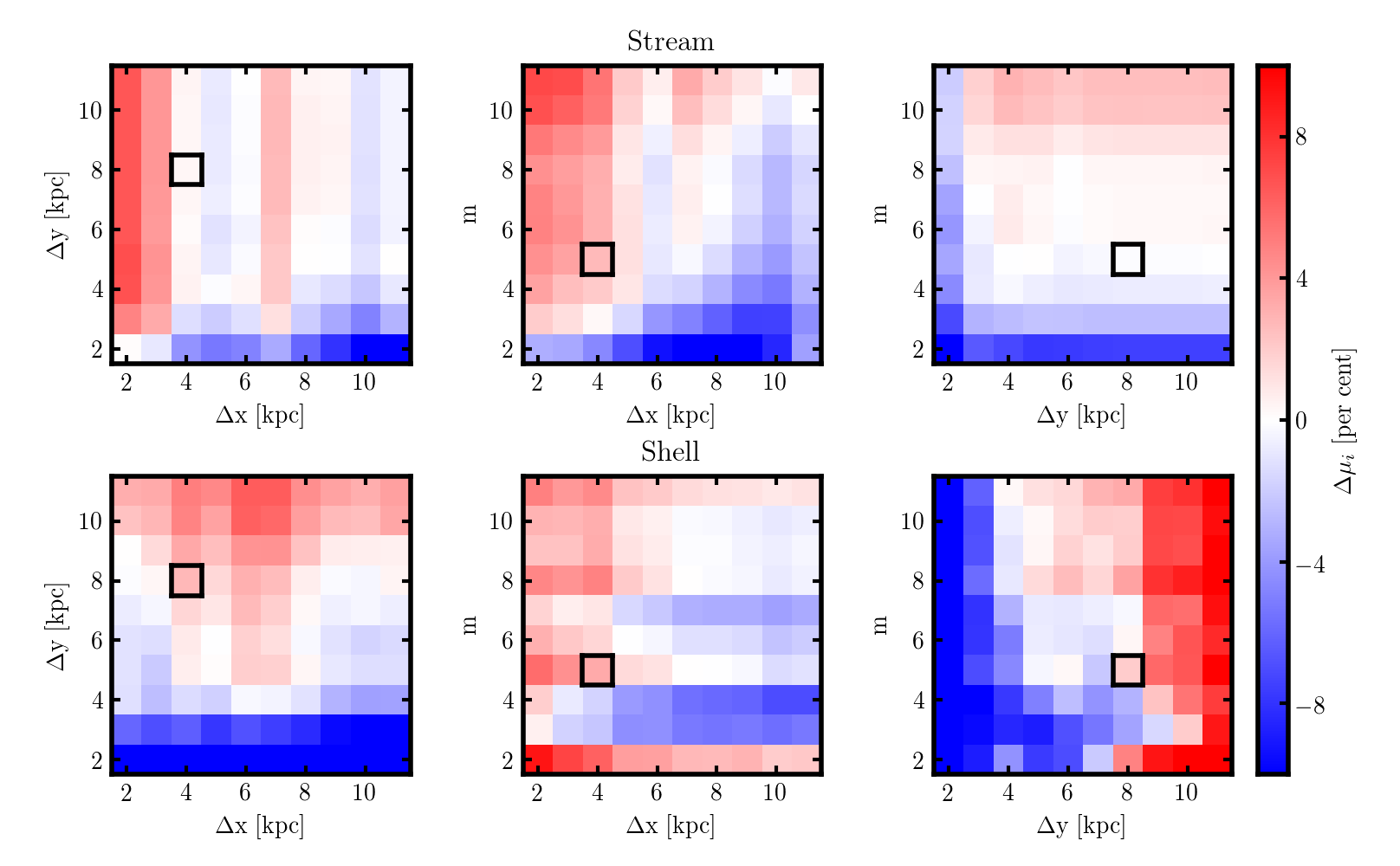}
        \caption{Variation of the local morphology metric $\mu_{i}$ as a function of its parameters $m$, $\Delta x$, and $\Delta y$, which correspond to the number of basis functions used and the area over which the expansion is computed. The top and bottom rows correspond to the same ridgepoints as shown in Figure~\ref{fig:pointclassifier}, representing locations on a stream and a shell, respectively. The quantity held constant is set to its fiducial value ($m=5$,  $\Delta x= 4$ kpc, and  $\Delta y = 8$ kpc) and these same values are highlighted by black boxes in each panel. Color indicates a percentage difference from the median of each panel. The computed value varies by only a few percent for a wide range of choices; the most obvious constraint is that the range $[y_{\mathrm{min}}, y_{\mathrm{max}}]$ should be at least twice the KDE smoothing scale so that the shell edge is captured correctly. This behavior is typical among sample points tested.}
    \label{fig:fft_variations}
\end{figure*}

A density ridge is defined by a set of conditions on both the gradient and the Hessian of the density field, which we estimate using a kernel density estimate (KDE): 
\begin{equation}
{\rho}(\vec{x}) = \frac{1}{nh^2}\sum_{i=1}^{n} K \left( \frac{|\vec{x}-\vec{X}_i|}{h} \right)
\end{equation}
where $n$ is the number of data points, $K$ is a kernel (in this work, the Gaussian kernel), and $h$ is a smoothing bandwidth. The choice of $h$ is an important but difficult issue. If the adopted value is too large, the KDE will over--smooth the debris and hamper classification by eliminating fine structure; too small, and the resulting density field will have too many isolated modes for the test points to effectively find the principal curves. A similar problem exists in identification of tidal debris in image data. In that case it is often found to be advantageous to smooth the data at a scale above the limit set by the point spread function; a kernel of width 1--3 kpc can enhance the low surface brightness features \citep[e.g.~][]{2011A&A...536A..66M, 2018ApJ...857..144H, 2018arXiv180403330M}. Here we adopt $h=2$ kpc.

Next, the gradient $g(\vec{x}) = \nabla \rho(\vec{x})$ and Hessian $H(\vec{x}) = \nabla \nabla \rho(\vec{x})$ are computed and $H(\vec{x})$ is eigendecomposed. Defining $\vec{v}(\vec{x})$ as the eigenvector corresponding to the smallest eigenvalue of $H$, the ridge is the set of points $\vec{x}_r$ that satisfy
\begin{equation} \label{eq:constr}
\vec{v}(\vec{x}_r)^{\transpose} \nabla \rho(\vec{x}_r) = 0 \ \mathrm{and} \ \vec{v}(\vec{x}_r)^{\transpose} H(\vec{x}_r)\vec{v}(\vec{x}_r) < 0
\end{equation}
\citep{eberly_ridges_1996, SCMS, genovese2014, chen2015}. These relations demand that the density field has a local maximum in the eigendirection corresponding to the smallest eigenvalue of $H$. The eigendirection corresponding to the larger eigenvalue is also the direction of the gradient and therefore points along the ridgeline; we note this direction $\vec{\phi}_r$ for each $\vec{x}_r$ for use in further analysis. Examples of the principal curves that result from this SCMS procedure are shown in Figure~\ref{fig:scms_example}. In the top panel,  the ridgeline cleanly traces the center of the tidal stream, even capturing the `s--shape' characteristic of the progenitor position where stars are leaving from the inner and outer Lagrange points to form the tidal tails. The shell debris on the bottom panel is more complex; the shell caustics are clearly marked out but also traced are the `pericenter streams' where the particles pass through their host as they move along their orbits from one shell to another. There are also some scattered ridgeline points that represent locations where the smoothing bandwidth is insufficient to connect with major structures; in this snapshot about 1\% of the final positions are spurious for this reason.

\begin{figure*}
	\includegraphics[width=\linewidth]{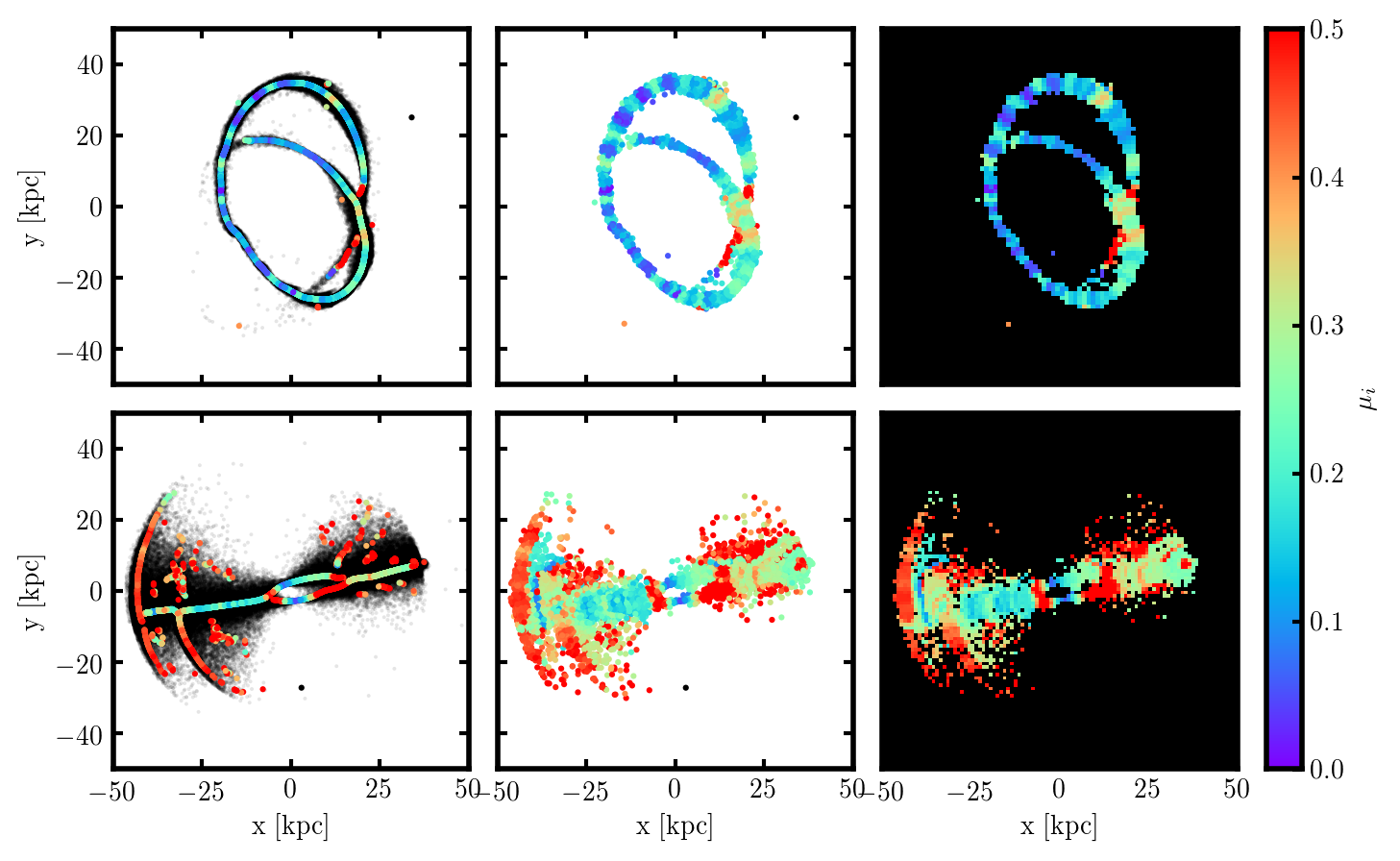}
        \caption{Global morphological classification for simulation snapshots of a stream (top) and shell (bottom). Left: the local morphology $\mu_{i}$ is evaluated for each ridge point. One can immediately see that the stream is more symmetric, with only a few points having less than $\sim 80\%$ of the orthogonal series expansion's density represented in symmetric terms. This gives the stream $\mu_{\mathrm{S}} = 0.17$. On the other hand, ridgepoints in the shell simulation have a median value $\mu_{\mathrm{S}} = 0.32$, representing a typical asymmetry nearly twice that of the stream simulation. Center: the initial positions $\vec{y}_i$ of the ridge points, color--coded by $\mu_{i}$ calculated at their final positions. Right: the mean value of the average morphology of gridded cells gives an area-weighted view of the morphology.}
    \label{fig:globalmorph}
\end{figure*}

\subsubsection{Feature classification: orthogonal series density estimation}

With the ridge points identified as locations of interest, the next step is to examine the structure of the particle density in the vicinity to determine if the ridgeline corresponds to a shell--like or stream--like feature. The key morphological identifier of shells is a gradual rise in surface density with increasing radius followed by a sharp drop to nearly zero; the density is highly asymmetric perpendicular to the shell edge (which the ridgeline parallels). Streams, on the other hand, tend to be more symmetric about their principal curve. We therefore take as an ansatz that each ridge point $\vec{x}_r$ can be classified on the basis of this asymmetry in the direction perpendicular to $\vec{\phi}_r$, i.e.~perpendicular to the ridgeline. We investigated a number of ways to characterize the symmetry of the density field including distribution properties like skewness and kurtosis, the Wilcoxon signed-rank test \citep{wilcoxon}, the response of a Sobel filter \citep{sobel}, and non--parametric mixture models \citep{patra2016}. While each of these tools have some advantages, extensive testing suggests that a decomposition based on a Fourier series provides conceptual simplicity, high speed, and an effective classification. 

Having chosen this tool, the SCUDS methodology for classifying each ridge point is illustrated in Figure~\ref{fig:pointclassifier}. First, a subset of particles in the vicinity of $\vec{x}_r$ on which to operate is constructed. The local principal curve is shifted to the origin and axis--aligned by rotating the data by $\vec{\phi}_r$ such that the ridgeline lies along the $x$ axis. Particles in the range $y_{\mathrm{min}}<y<y_{\mathrm{max}}$ and $x_{\mathrm{min}}<x<x_{\mathrm{max}}$ are selected and their $x$ coordinate is discarded, producing a one-dimensional particle distribution perpendicular to the ridgeline. 
Next, an {\em orthogonal series density estimator} is applied. The idea is to generate an approximation $\hat{f}$ of the true density $f$ by estimating its Fourier expansion coefficients using samples drawn from it. Here the samples are the simulation particle's positions in this locally ridge--aligned, one--dimensional coordinate system. More explicitly, after a change of variable such that $(y_{\mathrm{min}}, y_{\mathrm{max}})\rightarrow (-\pi,\pi)$, the integrals for calculating the coefficients $a_j$ and $b_j$

\begin{equation}
a_j = \frac{1}{\pi}\int_{-\pi}^\pi f(x) \cos jx\ dx,\ 
b_j = \frac{1}{\pi}\int_{-\pi}^\pi f(x) \sin jx\ dx
\end{equation}

\noindent are approximated by using a sum of Dirac delta functions at the particle positions to represent the density, $f(x) = \sum_k \delta(x-x_k)$, where $k$ is over all selected particles. The estimated coefficients are therefore 

\begin{equation}\label{eq:coeff}
\hat{a}_j = \frac{1}{\pi}\sum_k \cos jx_k, \ 
\hat{b}_j = \frac{1}{\pi}\sum_k \sin jx_k,
\end{equation}

\noindent and the orthogonal series density estimator is 
\begin{equation} \label{eq:estimator}
\hat{f}(x; m) = \frac{N}{2\pi}\left[1 + 2\sum_1^m (\hat{a}_j \cos{jx} + \hat{b}_j \sin{jx}) \right]
\end{equation}

\noindent where $m$ is the maximum basis function degree of interest; this corresponds to an effective smoothing, since as $m\rightarrow \infty$ the series must converge to the set of delta functions it uses as input, not $f$. 

A measure of the total degree of asymmetry in $\hat{f}$ can be computed by comparing the sine and cosine contributions to the overall density; we define the local morphology metric as
\begin{equation}\label{eq:mui}
\mu_{i} \equiv \frac{\sum_{j=1}^{m} |\hat{b}_j|}{\sum_{j=1}^{m} |\hat{a}_j|+|\hat{b}_j|},
\end{equation}
\noindent which is computed for each ridge point and captures the scale of the local density asymmetry.
Note that the $j=0$ term that represents the mean value is not included. We presume that any global asymmetry due to the smooth stellar halo, which has typical scales much greater than the smoothing length considered here, can be removed by fitting a power law -- perhaps after masking the regions of interest identified by SCUDS.

\subsubsection{Parameter choices}

The SCUDS scheme defined above has introduced a number of free parameters: the scale of the kernel density estimate $h$, the maximum Fourier term $m$ (which corresponds to an effective smoothing) and the data selection ranges $[x_{\mathrm{min}},x_{\mathrm{max}}]$ and $[y_{\mathrm{min}},y_{\mathrm{max}}]$. Ideally the classifier will be insensitive to the detailed choice of each parameter. To test this, we again examined the same test point as in Figure~\ref{fig:pointclassifier} and varied the classification parameters. The effect this has on $\mu_{i}$ is show in Figure~\ref{fig:fft_variations}; the top row is the point on the stream, and the bottom is on the shell edge. The color indicates the percentage variation for each choice relative to the median value in the panel. One can see that $\mu_i$ varies by only a few percent over a wide range of choices for each parameter, with the exception of $y_{\mathrm{min}}-y_{\mathrm{max}}$. When this quantity is less than about twice the KDE smoothing scale the shape of the density histogram qualitatively changes since the shell edge is not included in the range. Very large box sizes also cause difficulties as the density expansion has to capture multiple features. For the remainder of this work we set $h=2$ kpc, $m = 5$, $x_{\mathrm{max}}-x_{\mathrm{min}} = 2h$, and $y_{\mathrm{max}}-y_{\mathrm{min}} = 4h$.

\begin{figure*}
\subfloat{\includegraphics[width=\linewidth]{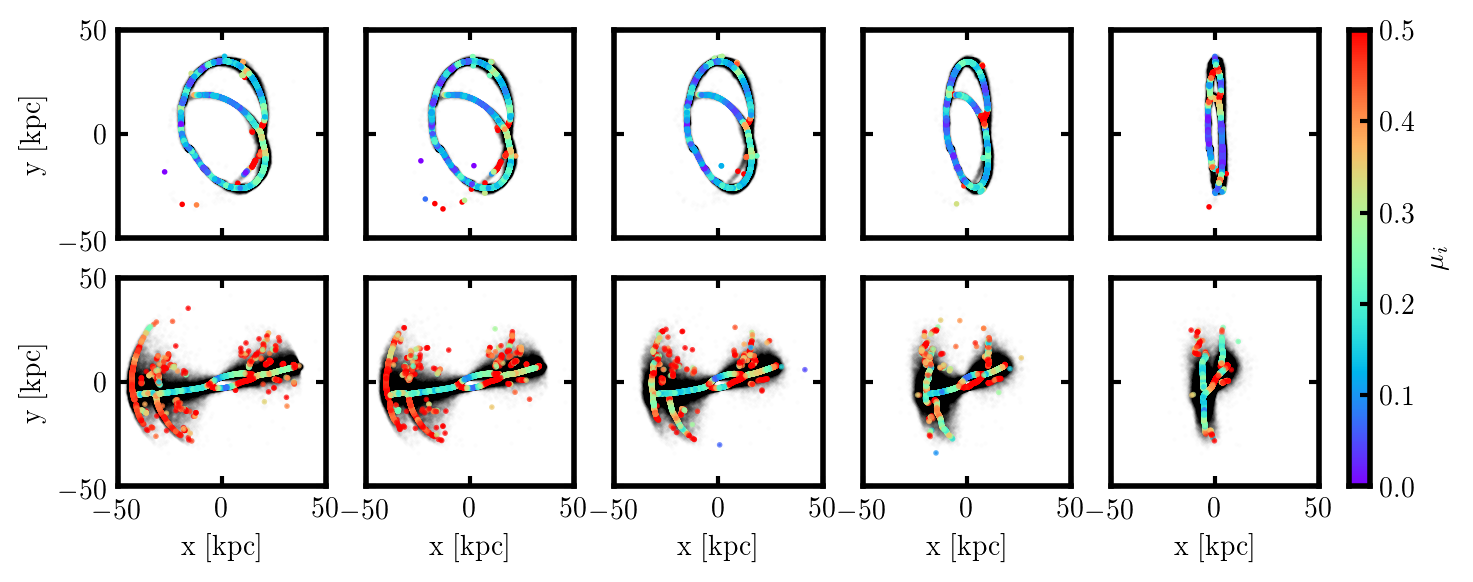}}\\
\subfloat{\includegraphics[width=\linewidth]{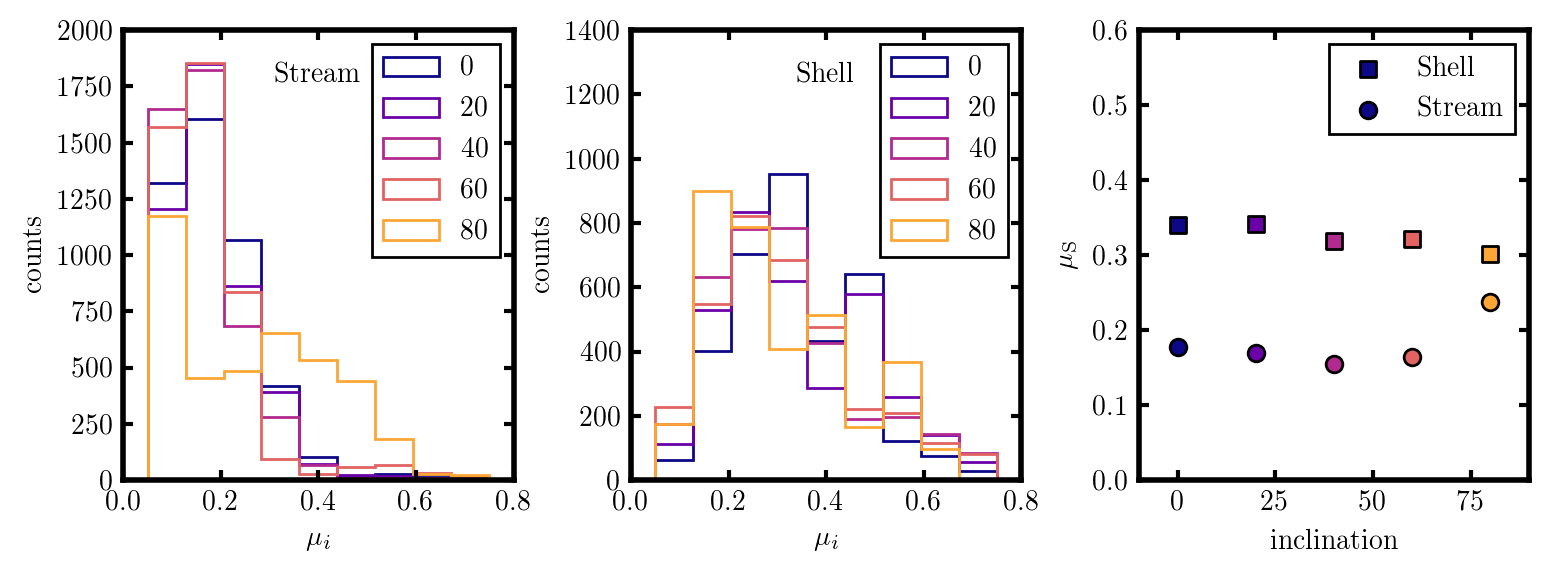}}
\caption{Effect of inclination on the ridgepoint and global morphology. Top row: the same simulation as the stream simulation (top row) of Figure~\ref{fig:globalmorph}, except inclined at different angles to the line of sight. The morphology is calculated in the same way. Second row: same as top row, but for the shell simulation (bottom row of Figure~\ref{fig:globalmorph}). Bottom left and bottom center: distribution of point morphologies $\mu_{i}$ for the inclined stream and shell simulations, respectively. Right: mean morphology $\mu_{\rm S}$ calculated as a function of inclination; the automatic classifier is well behaved for a wide range of inclinations. 	\label{fig:incl}}
\end{figure*}

\subsubsection{Source reconstruction \& debris classification}

Given the above method for analyzing each ridgepoint, next an overall morphology for each simulation snapshot must be determined. We use the simplest possible method: computing the median value of the local morphology $\mu_{i}$ (Equation~\ref{eq:mui}) over all ridgepoints. This gives in some sense a number--density--weighted morphology since the initial SCMS positions are chosen as a subset of the data, $\vec{y} \in \{X_i\}$. We denote this global morphology ${\mu_{\rm S}}$. Although the original goal was to replicate automatically the shell versus stream decisions of human classifiers, this continuous variable classification approach allows more insight into the evolutionary state of the tidal debris, as shown below.


While in principle one could initialize the ridge initial positions as every particle, in practice this is rather computationally expensive and, we find, unnecessary. To ensure this method can return consistent morphology results, we computed $\mu_{\mathrm{S}}$ using 100 random initializations of 500 particles from a simulation snapshot. The standard deviation the median was only $0.6\%$ of the mean, which we consider negligible. To err on the side of caution, for the remainder of this work we use 5,000 ridge particles.

Finally, stars from the host galaxy will make any kind of analysis difficult near its center; for example, \cite{2018arXiv180505970K} found that their detection efficiency dropped dramatically within $\sim 4 R_e$. With {\em WFIRST} in mind and noting that in the mid--infrared nearby early--type galaxies have effective radii $\sim 3$ kpc \citep{2017MNRAS.464.4611F}, we remove any ridgepoints within 15 kpc of the host's center. Similarly, if a bound remnant exists any ridgepoints within 5 kpc of it are masked.

\section{Results}
\label{sec:results}

Having defined a method for the automatic identification and classification of tidal debris, we now explore several applications. The SCUDS morphology metric $\mu_S$ has been tabulated for each of the simulations described above and is provided along with the associated code in a public repository\footnote{https://github.com/davidhendel/scuds}.

\subsection{Inclination effects}

One difficult issue in by--eye detection and morphological classification is the effect of inclination. Using the tool described above this confounder can be addressed directly, both in terms of changes to the appearance of specific points on the debris and the global morphology. 

Figure~\ref{fig:incl} explores this for two representative simulations, again one shell and one stream. Perhaps unsurprisingly, the relatively one-dimensional shape of the stream is unaffected by rotation until the inclination $i \gtrsim 75^\circ$, where projection causes substantial overlap between the leading and trailing streams. The slight decrease at small inclinations is due to the line--of--sight projection hiding epicyclic overdensities or `feathers' as they are superimposed on top of the main arm.

On the other hand, the sharp density drop at the edge of shells is enhanced by projection, in the same manner as limb--brightening, and one might expect that as the densest part of the shell (in the orbital plane) is rotated into the line of sight that its appearance would change significantly. This is borne out as shown in Figure~\ref{fig:incl}; the automatic classifier still detects the shell edge, but due to decreasing asymmetry its morphology $\mu_{\rm S}$ decreases from $\sim 0.33$ to $\sim 0.29$ for inclinations of 0 and 80 degrees, respectively. The simulations are necessarily symmetric when viewed edge--on, due to the symmetry of the spherical potential and initial conditions, so an overall decrease is unsurprising. However, it seems that the SCUDS morphology is relatively insensitive to inclination effects.  

\subsection{Time evolution}

\begin{figure}
	\includegraphics[width=\linewidth]{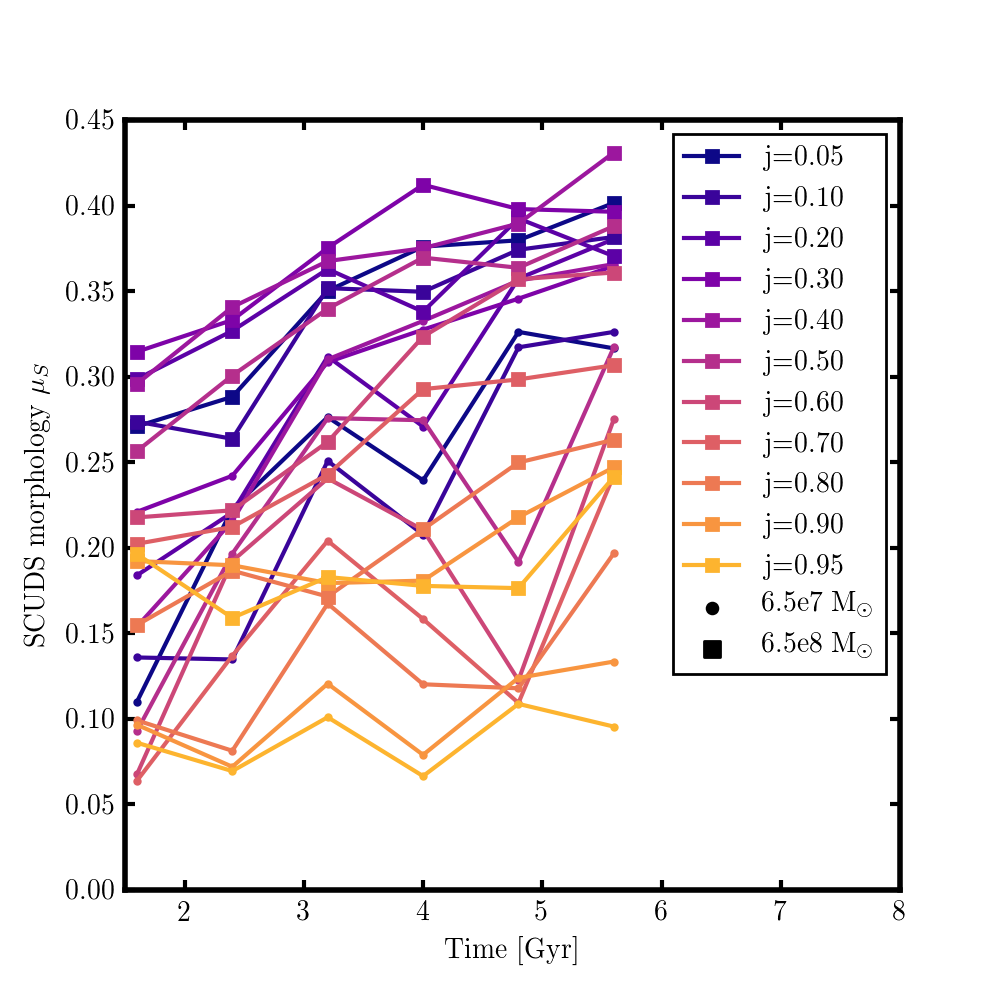}
        \caption{Time evolution of the SCUDS morphology metric $\mu_S$ for satellites of mass $m=6.5\times10^7$ and $m=6.5\times10^8$ M$_\odot$ on a variety of orbits with circularities $j=L/L_{\rm circ}$ between 0.05 and 0.95. Some trends are visible: the more massive satellites (large squares) tend to have a larger value of $\mu_S$, while the lower mass satellites (small circles) on very radial orbits tend to have a morphology that increases more quickly with time. Relatively circular orbits tend to remain at small values for $\mu_S$ for the duration of the simulation.}
    \label{fig:time}
\end{figure}

\begin{figure}
	\includegraphics[width=\linewidth]{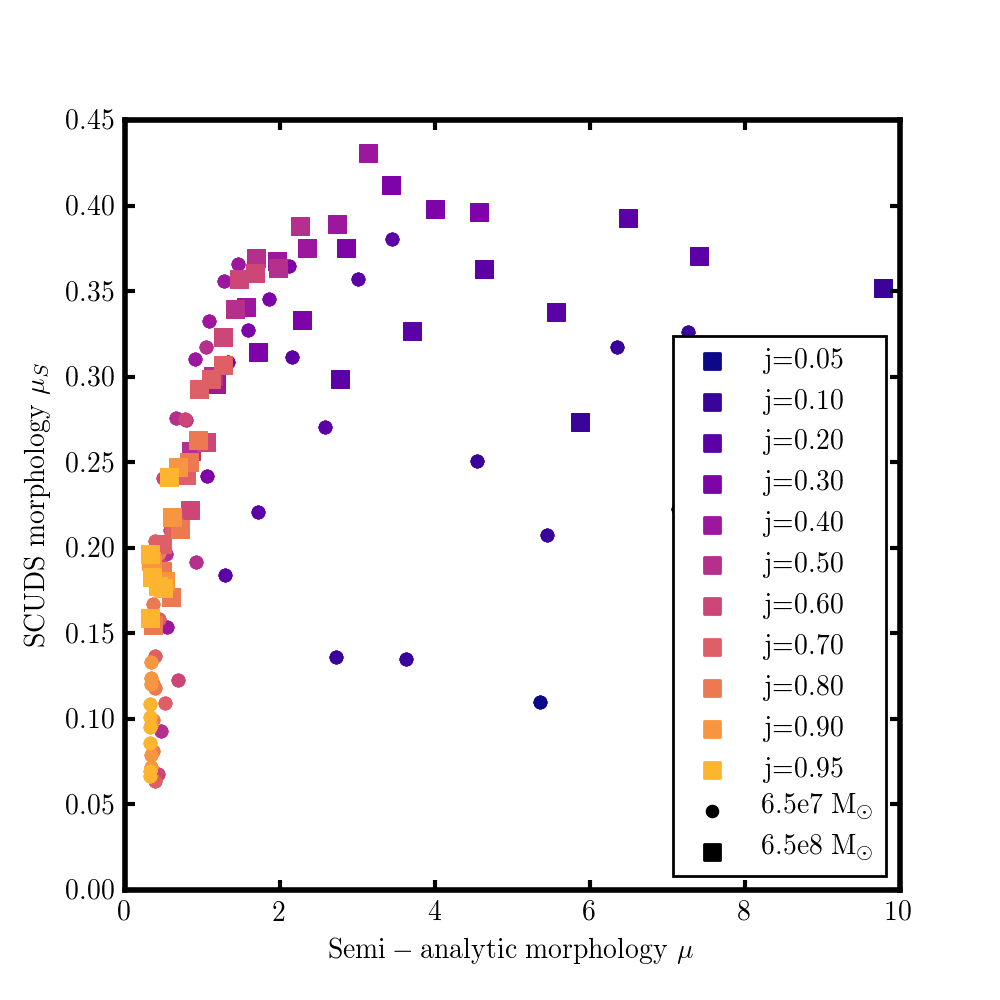}
        \caption{A comparison of the SCUDS morphology metric $\mu_S$ with the semi--analytic morphology metric $\mu$ for the same simulation snapshots as in Figure~\ref{fig:time}. Stream--like debris has $\mu \lesssim 1$; we find that snapshots in this regime also have $\mu_S \lesssim 0.275$, and the two are highly correlated. The few points with $\mu_S < 0.275$ and $\mu > 1$ are due to debris that is in the `pericenter stream' state but which has not mixed in radial phase enough to also fill out shells during this time.}
    \label{fig:morph_comp}
\end{figure}

A clear expectation from a variety of analytic arguments \citep{2015MNRAS.450..575A, 2015MNRAS.454.2472H}, cosmologically motivated restricted simulations \citep{2008ApJ...689..936J}, and fully cosmological hydrodynamic simulations \citep{2017arXiv170606102P} is that debris from satellites on orbits of intermediate circularity should slowly transition from stream--like to shell--like over time. This can easily be evaluated by SCUDS, as seen in Figure~\ref{fig:time}. There is a clear pattern for many debris structures, especially from the more massive group, to slowly grow more asymmetric (i.e.~have more of their mass in shells) as time goes on. For the high--mass satellites on the most radial orbits, which generate shell structures almost immediately, this progress is approximately linear in time which also agrees with the analytic expectation (see Section~\ref{sec:mm}).

On the other hand, the low--mass satellites on relatively circular orbits have little evolution away from their original stream--like state over the more than 5 Gyr shown here. In this case, the primary source of snapshot--to--snapshot variation is due to the epicyclic `feathers' that form parts of the stream \citep[e.g.~][]{2010MNRAS.401..105K} which vary in prominence depending on a number of factors including the amount of recent mass loss and current orbital phase. Several of these structures are visible in the top left panel of Figure~\ref{fig:pointclassifier} and in particular the one inside the selection box at $\Delta$y $\sim$2 produces some of the slight asymmetry seen in the corresponding density histogram and orthogonal series density estimator. Occasionally these streams see a sudden increase in asymmetry on the order of tens of percent; this corresponds to the first snapshot in which the leading and trailing tails cross, which unsurprisingly introduces significant asymmetry in that region. The crossing may or may not persist as the simulation continues.

\subsection{Connection to the morphology metric}
\label{sec:mm}

Here we briefly review the computation of the morphology metric of \cite{2015MNRAS.454.2472H}; see that work for a complete justification. Given an orbit in the host potential with energy $E_{\rm orb}$ and angular momentum $L_{\rm orb}$, the angular size of the shell edge due to azimuthal precession is just the differential precession $\Delta \psi$ per orbit with respect to $L$ at $E_{orb}$ times the angular momentum scale and the number of orbits $N_{\rm orb}$: 
\begin{equation}\label{eq:psil}
\Psi_L \equiv l_s N_{\rm orb} \left.\frac{\partial \Delta \psi}{\partial L}\right|_{E_{orb}}; l_s = \sigma R_p + 2s V_p R_p = (\sqrt{3} + 2) sL
\end{equation}
\noindent where the contribution proportional to $\partial \Delta \psi/\partial E$ is neglected due to the weak dependence of $\Delta \psi$ on $E$, $V_p$ and $R_p$ are the pericentric velocity and radius, respectively, and
\begin{equation}
s = \left( \frac{m}{3 M(R_p)}\right) ^ {1/3} 
\end{equation}
\noindent where $M(r)$ denotes the host's mass enclosed at radius $r$.

Similarly, the length of the stream along its orbit can be approximated by considering how far the fastest and slowest unbound stars have had time to move; this is given by
\begin{equation}\label{eq:psie}
\Psi_E \equiv e_s N_{\rm orb}\left.\frac{\Delta \psi}{T_r}\frac{\partial T_r}{\partial E}\right|_{L_{\rm orb}}; e_s = 2sR_p\left.\frac{\partial \Phi}{\partial R}\right|_{R_p}
\end{equation}
\noindent where $T_r$ indicates the radial period. The value of $\Psi_E$ is bounded by the width of the orbit's rosette petals, denoted $\alpha$ and defined as the position angle traveled in the during the half--period closest to apogalacticon. Finally, the morphology metric is calculated as

\begin{equation}
\mu \equiv \frac{\Psi_L}{\mathrm{min}\left(\alpha, \Psi_E \right)}.
\end{equation}

\noindent Previously, it was shown that $\mu = 1$ provided an excellent boundary between debris that would be visually classified as shells (with $\mu$ greater than 1) and streams ($\mu$ less than 1). 

Figure~\ref{fig:morph_comp} compares the semi--analytic morphology metric $\mu$ with the results of the SCUDS algorithm. For a given simulation the two values are tightly correlated, with few exceptions. In particular, from inspection of Figure~\ref{fig:morph_comp}, the simulation space that has $\mu<1$ also has $\mu_{\rm S} \lesssim 0.275$. At larger values of $\mu$ there is more scatter, driven by the most eccentric orbits of the lower--mass satellite. Inspection of these outliers reveals that the orbital period differences in this simulation are so small that it does not completely fill an orbit from apocenter to apocenter, and so it transitions from looking like a stream near pericenter to a shell near apocenter depending on where exactly the snapshot is taken. We conclude that $\mu_{\rm S}$ is an excellent proxy for $\mu$.

\section{Discussion}
\label{sec:discussion}

At first glance this correspondence between the semi--analytic morphology estimate $\mu$ which requires complete knowledge of the interaction parameters and the asymmetry estimate $\mu_S$ seems prosaic, but \cite{2015MNRAS.454.2472H} showed that, given a particular cosmology resulting in specific merger rates, mass--concentration relations, and so on, the orbital infall distribution determines what types of debris structures should be observed, i.e.~what the expected distribution of $\mu$ will be. The correspondence between  $\mu_{\rm S}$ and $\mu$ therefore gives us some hope of inverting this relation and determining the orbits of destroyed satellites based on their tidal debris in a statistical sense. One can imagine comparing a synthetic survey built from cosmological N--body simulations to observations using a tool like SCUDS. One distinct limitation so far is that we have considered only relatively minor mergers; while these are the most common, they may not be the most frequently observed. As we have demonstrated it is straightforward to investigate this type of question with SCUDS using an appropriate dataset.

This is far from the end point of possible algorithmic development. So far we have used only a fraction of the available information from either the Subspace--Constrained Mean Shift output or the orthogonal density expansion. As an example, the simplest possible addition would be to check the local density in the vicinity of each ridge point. Using a minimum density threshold might allow the removal of spurious ridgepoint locations where the particle density was too low for the adopted KDE smoothing allow free movement towards the principal curves. The information required for such a test is already available in the kernel density estimate.

In addition, we expect that `real' parts of debris that have meaningful classifications should produce a coherent signal in the direction field $\vec{\phi}_r$. SCUDS could be extended to check if an individual ridge point was excessively misaligned with others near its position. This may be especially helpful in identifying areas where e.g.~the leading and trailing stream cross due to phase mixing or otherwise overlap when projected on the sky, a typical source of excess asymmetric signal.

Another possibility is to more carefully examine the density expansion. Not all basis functions are equally informative about the overall distribution and low signal--to--noise coefficients may skew the results, particularly in the low density regime. One way to evaluate this is to examine the basis functions' second moment matrix;
\cite{1996ApJ...470..715W} showed that the trace of this matrix is related to the signal--to--noise ratio in each basis function. One could imagine using this property to either optimize $m$, the maximum term  used in the calculation of $\mu_{\rm S}$, or to estimate its uncertainty.

SCUDS has been designed to work on particles with the intention of approximating resolved stellar populations data. This type of data is powerful but difficult to acquire; currently single--digit numbers of galaxies have been explored in this way, and large samples will have to wait until the next generation of wide--field space--based observatories. Extension of this algorithm to operate on imaging data would open up the realm of possible applications enormously, since there are a number of statistically interesting datasets currently available such as those of \cite{2013ApJ...765...28A, 2018ApJ...857..144H}, and \cite{2018arXiv180505970K}. The primary algorithmic difference will be in calculating the derivatives required to evaluate the Hessian in Equation~\ref{eq:constr} as well as sensibly resampling the pixel data to obtain the density structure perpendicular to the principal curves. We are confident that both of these challenges can be overcome.

In addition to illuminating the orbits of satellite galaxies, shell systems in particular have been proposed as a way to attack several crucial problems in galactic dynamics. One of these is to determine the total mass distribution of galaxies, which is of particular importance beyond the limits of optical or neutral hydrogen rotation curves. The phase space caustic of an individual shell edge is a probe of the local gravitational force and so possible way forward is to use shell systems to put constraints on the host's mass distribution \citep{2012A&A...545A..33E, 2013MNRAS.435..378S}. Such methods could also benefit from the precise, objective localization of the configuration space structure provided by SCUDS.

\section{Conclusion}
\label{sec:conclusions}

In this work we have developed a new and fully automated way to identify and classify substructure, in particular tidal debris structures from disrupting satellite galaxies, using a machine--vision method that we dub Subspace Constrained Unsupervised Detection of Structure, or SCUDS. The two basic components of this algorithm are Subspace--Constrained Mean Shift, which is used to identify the `ridgelines' of high density that trace the tidal features, in combination with an orthogonal series density estimator that is used to describe the structure perpendicular to the ridgelines. Whether this structure is symmetric or asymmetric is used locally classify the parts of the ridge that appear like shells or like streams.

We demonstrate that this tool can effectively deal with common observational issues, for example inclination effects, and that it is insensitive to many of the parameter choices that are made. Most importantly, we illustrate a strong correlation between the SCUDS morphological indicator and a semi--analytic one that is known to predict the correct debris morphologies in this data set. This means that the SCUDS classifications are interpretable in terms of the interaction parameters such as the host--to--satellite mass ratio, the interaction time, and the satellite's orbit. We also discuss several possible extensions that will make it even more robust and informative. Tools like SCUDS are necessary to understand the truly large datasets that will flow from near--future surveys and will provide a clearer view of galactic assembly.   

\section*{Acknowledgements}

The authors thank Robyn Sanderson for useful discussions and Rahul Biswas for work on the initial stages of this project. DH and KVJ acknowledge support from NASA through subcontract JPL 1558281 and ATP grant NNX15AK78G, as well as from the NSF through the grant AST-1614743. DH also acknowledges financial support from the Natural Sciences and Engineering Research Council of Canada (NSERC; funding reference number RGPIN-2015-05235) and an Ontario Early Researcher Award (ER16-12-061). We acknowledge the use of computing resources from Columbia University's Shared Research Computing Facility project, which is supported by NIH Research Facility Improvement Grant 1G20RR030893-01 and associated funds from the New York State Empire State Development Division of Science Technology and Innovation (NYSTAR) Contract C090171. This work made use of SciPy \citep{scipy}, Matplotlib \citep{Hunter:2007}, helit \citep{helit}, IPython \citep{ipython}, Astropy \citep{2013A&A...558A..33A, 2018arXiv180102634T}, and Numba \citep{numba}.

%


\bibliographystyle{mnras}
\bibliography{mv.bib,nonadsabsrefs.bib} 

\begin{thebibliography}{}
\makeatletter
\relax
\def\mn@urlcharsother{\let\do\@makeother \do\$\do\&\do\#\do\^\do\_\do\%\do\~}
\def\mn@doi{\begingroup\mn@urlcharsother \@ifnextchar [ {\mn@doi@}
  {\mn@doi@[]}}
\def\mn@doi@[#1]#2{\def\@tempa{#1}\ifx\@tempa\@empty \href
  {http://dx.doi.org/#2} {doi:#2}\else \href {http://dx.doi.org/#2} {#1}\fi
  \endgroup}
\def\mn@eprint#1#2{\mn@eprint@#1:#2::\@nil}
\def\mn@eprint@arXiv#1{\href {http://arxiv.org/abs/#1} {{\tt arXiv:#1}}}
\def\mn@eprint@dblp#1{\href {http://dblp.uni-trier.de/rec/bibtex/#1.xml}
  {dblp:#1}}
\def\mn@eprint@#1:#2:#3:#4\@nil{\def\@tempa {#1}\def\@tempb {#2}\def\@tempc
  {#3}\ifx \@tempc \@empty \let \@tempc \@tempb \let \@tempb \@tempa \fi \ifx
  \@tempb \@empty \def\@tempb {arXiv}\fi \@ifundefined
  {mn@eprint@\@tempb}{\@tempb:\@tempc}{\expandafter \expandafter \csname
  mn@eprint@\@tempb\endcsname \expandafter{\@tempc}}}

\bibitem[\protect\citeauthoryear{{Aihara} et~al.,}{{Aihara}
  et~al.}{2018}]{2018PASJ...70S...4A}
{Aihara} H.,  et~al., 2018, \mn@doi [Publications of the Astronomical Society
  of Japan] {10.1093/pasj/psx066}, \href
  {https://ui.adsabs.harvard.edu/#abs/2018PASJ...70S...4A} {70, S4}

\bibitem[\protect\citeauthoryear{{Amorisco}}{{Amorisco}}{2015}]{2015MNRAS.450..575A}
{Amorisco} N.~C.,  2015, \mn@doi [\mnras] {10.1093/mnras/stv648}, \href
  {https://ui.adsabs.harvard.edu/#abs/2015MNRAS.450..575A} {450, 575}

\bibitem[\protect\citeauthoryear{{Astropy Collaboration} et~al.,}{{Astropy
  Collaboration} et~al.}{2013}]{2013A&A...558A..33A}
{Astropy Collaboration} et~al., 2013, \mn@doi [\aap]
  {10.1051/0004-6361/201322068}, \href
  {https://ui.adsabs.harvard.edu/#abs/2013A&A...558A..33A} {558, A33}

\bibitem[\protect\citeauthoryear{{Atkinson}, {Abraham}  \&
  {Ferguson}}{{Atkinson} et~al.}{2013}]{2013ApJ...765...28A}
{Atkinson} A.~M.,  {Abraham} R.~G.,   {Ferguson} A. M.~N.,  2013, \mn@doi
  [\apj] {10.1088/0004-637X/765/1/28}, \href
  {https://ui.adsabs.harvard.edu/#abs/2013ApJ...765...28A} {765, 28}

\bibitem[\protect\citeauthoryear{{Bailin}, {Bell}, {Chappell}, {Radburn-Smith}
  \& {de Jong}}{{Bailin} et~al.}{2011}]{2011ApJ...736...24B}
{Bailin} J.,  {Bell} E.~F.,  {Chappell} S.~N.,  {Radburn-Smith} D.~J.,   {de
  Jong} R.~S.,  2011, \mn@doi [\apj] {10.1088/0004-637X/736/1/24}, \href
  {https://ui.adsabs.harvard.edu/#abs/2011ApJ...736...24B} {736, 24}

\bibitem[\protect\citeauthoryear{{Barker}, {Ferguson}, {Irwin}, {Arimoto}  \&
  {Jablonka}}{{Barker} et~al.}{2012}]{2012MNRAS.419.1489B}
{Barker} M.~K.,  {Ferguson} A. M.~N.,  {Irwin} M.~J.,  {Arimoto} N.,
  {Jablonka} P.,  2012, \mn@doi [\mnras] {10.1111/j.1365-2966.2011.19814.x},
  \href {https://ui.adsabs.harvard.edu/#abs/2012MNRAS.419.1489B} {419, 1489}

\bibitem[\protect\citeauthoryear{{Belokurov} et~al.,}{{Belokurov}
  et~al.}{2006}]{2006ApJ...642L.137B}
{Belokurov} V.,  et~al., 2006, \mn@doi [\apj] {10.1086/504797}, \href
  {https://ui.adsabs.harvard.edu/#abs/2006ApJ...642L.137B} {642, L137}

\bibitem[\protect\citeauthoryear{{Bullock} \& {Johnston}}{{Bullock} \&
  {Johnston}}{2005}]{2005ApJ...635..931B}
{Bullock} J.~S.,  {Johnston} K.~V.,  2005, \mn@doi [\apj] {10.1086/497422},
  \href {https://ui.adsabs.harvard.edu/#abs/2005ApJ...635..931B} {635, 931}

\bibitem[\protect\citeauthoryear{{Chang}, {Macci{\`o}}  \& {Kang}}{{Chang}
  et~al.}{2013}]{2013MNRAS.431.3533C}
{Chang} J.,  {Macci{\`o}} A.~V.,   {Kang} X.,  2013, \mn@doi [\mnras]
  {10.1093/mnras/stt434}, \href
  {https://ui.adsabs.harvard.edu/#abs/2013MNRAS.431.3533C} {431, 3533}

\bibitem[\protect\citeauthoryear{Chen, Genovese  \& Wasserman}{Chen
  et~al.}{2015a}]{chen2015}
Chen Y.-C.,  Genovese C.~R.,   Wasserman L.,  2015a, \mn@doi [Ann. Statist.]
  {10.1214/15-AOS1329}, 43, 1896

\bibitem[\protect\citeauthoryear{{Chen}, {Ho}, {Freeman}, {Genovese}  \&
  {Wasserman}}{{Chen} et~al.}{2015b}]{2015MNRAS.454.1140C}
{Chen} Y.-C.,  {Ho} S.,  {Freeman} P.~E.,  {Genovese} C.~R.,   {Wasserman} L.,
  2015b, \mn@doi [\mnras] {10.1093/mnras/stv1996}, \href
  {https://ui.adsabs.harvard.edu/#abs/2015MNRAS.454.1140C} {454, 1140}

\bibitem[\protect\citeauthoryear{{Chen} et~al.,}{{Chen}
  et~al.}{2015c}]{2015MNRAS.454.3341C}
{Chen} Y.-C.,  et~al., 2015c, \mn@doi [\mnras] {10.1093/mnras/stv2260}, \href
  {https://ui.adsabs.harvard.edu/#abs/2015MNRAS.454.3341C} {454, 3341}

\bibitem[\protect\citeauthoryear{{Chen}, {Ho}, {Brinkmann}, {Freeman},
  {Genovese}, {Schneider}  \& {Wasserman}}{{Chen}
  et~al.}{2016}]{2016MNRAS.461.3896C}
{Chen} Y.-C.,  {Ho} S.,  {Brinkmann} J.,  {Freeman} P.~E.,  {Genovese} C.~R.,
  {Schneider} D.~P.,   {Wasserman} L.,  2016, \mn@doi [\mnras]
  {10.1093/mnras/stw1554}, \href
  {https://ui.adsabs.harvard.edu/#abs/2016MNRAS.461.3896C} {461, 3896}

\bibitem[\protect\citeauthoryear{{Chen} et~al.,}{{Chen}
  et~al.}{2017}]{2017MNRAS.466.1880C}
{Chen} Y.-C.,  et~al., 2017, \mn@doi [\mnras] {10.1093/mnras/stw3127}, \href
  {https://ui.adsabs.harvard.edu/#abs/2017MNRAS.466.1880C} {466, 1880}

\bibitem[\protect\citeauthoryear{{Cooper} et~al.,}{{Cooper}
  et~al.}{2010}]{2010MNRAS.406..744C}
{Cooper} A.~P.,  et~al., 2010, \mn@doi [\mnras]
  {10.1111/j.1365-2966.2010.16740.x}, \href
  {https://ui.adsabs.harvard.edu/#abs/2010MNRAS.406..744C} {406, 744}

\bibitem[\protect\citeauthoryear{{Crnojevi{\'c}} et~al.,}{{Crnojevi{\'c}}
  et~al.}{2016}]{2016ApJ...823...19C}
{Crnojevi{\'c}} D.,  et~al., 2016, \mn@doi [\apj] {10.3847/0004-637X/823/1/19},
  \href {https://ui.adsabs.harvard.edu/#abs/2016ApJ...823...19C} {823, 19}

\bibitem[\protect\citeauthoryear{{Dalcanton} et~al.,}{{Dalcanton}
  et~al.}{2009}]{2009ApJS..183...67D}
{Dalcanton} J.~J.,  et~al., 2009, \mn@doi [The Astrophysical Journal Supplement
  Series] {10.1088/0067-0049/183/1/67}, \href
  {https://ui.adsabs.harvard.edu/#abs/2009ApJS..183...67D} {183, 67}

\bibitem[\protect\citeauthoryear{{Duc} et~al.,}{{Duc}
  et~al.}{2015}]{2015MNRAS.446..120D}
{Duc} P.-A.,  et~al., 2015, \mn@doi [\mnras] {10.1093/mnras/stu2019}, \href
  {https://ui.adsabs.harvard.edu/#abs/2015MNRAS.446..120D} {446, 120}

\bibitem[\protect\citeauthoryear{Eberly}{Eberly}{1996}]{eberly_ridges_1996}
Eberly D.,  1996, Ridges in Image and Data Analysis, 1st edn.
Springer

\bibitem[\protect\citeauthoryear{{Ebrov{\'a}}, {J{\'\i}lkov{\'a}}, {Jungwiert},
  {K{\v{r}}{\'\i}{\v{z}}ek}, {B{\'\i}lek}, {Barto{\v{s}}kov{\'a}},
  {Skalick{\'a}}  \& {Stoklasov{\'a}}}{{Ebrov{\'a}}
  et~al.}{2012}]{2012A&A...545A..33E}
{Ebrov{\'a}} I.,  {J{\'\i}lkov{\'a}} L.,  {Jungwiert} B.,
  {K{\v{r}}{\'\i}{\v{z}}ek} M.,  {B{\'\i}lek} M.,  {Barto{\v{s}}kov{\'a}} K.,
  {Skalick{\'a}} T.,   {Stoklasov{\'a}} I.,  2012, \mn@doi [\aap]
  {10.1051/0004-6361/201219940}, \href
  {https://ui.adsabs.harvard.edu/#abs/2012A&A...545A..33E} {545, A33}

\bibitem[\protect\citeauthoryear{{Ferguson}, {Irwin}, {Ibata}, {Lewis}  \&
  {Tanvir}}{{Ferguson} et~al.}{2002}]{2002AJ....124.1452F}
{Ferguson} A. M.~N.,  {Irwin} M.~J.,  {Ibata} R.~A.,  {Lewis} G.~F.,   {Tanvir}
  N.~R.,  2002, \mn@doi [\aj] {10.1086/342019}, \href
  {https://ui.adsabs.harvard.edu/#abs/2002AJ....124.1452F} {124, 1452}

\bibitem[\protect\citeauthoryear{{Forbes}, {Sinpetru}, {Savorgnan},
  {Romanowsky}, {Usher}  \& {Brodie}}{{Forbes}
  et~al.}{2017}]{2017MNRAS.464.4611F}
{Forbes} D.~A.,  {Sinpetru} L.,  {Savorgnan} G.,  {Romanowsky} A.~J.,  {Usher}
  C.,   {Brodie} J.,  2017, \mn@doi [\mnras] {10.1093/mnras/stw2604}, \href
  {https://ui.adsabs.harvard.edu/#abs/2017MNRAS.464.4611F} {464, 4611}

\bibitem[\protect\citeauthoryear{Fukunaga \& Hostetler}{Fukunaga \&
  Hostetler}{1975}]{meanshift}
Fukunaga K.,  Hostetler L.,  1975, \mn@doi [IEEE Transactions on Information
  Theory] {10.1109/TIT.1975.1055330}, 21, 32

\bibitem[\protect\citeauthoryear{Genovese, Perone-Pacifico, Verdinelli  \&
  Wasserman}{Genovese et~al.}{2014}]{genovese2014}
Genovese C.~R.,  Perone-Pacifico M.,  Verdinelli I.,   Wasserman L.,  2014,
  \mn@doi [Ann. Statist.] {10.1214/14-AOS1218}, 42, 1511

\bibitem[\protect\citeauthoryear{{Greco} et~al.,}{{Greco}
  et~al.}{2017}]{2017arXiv170904474G}
{Greco} J.~P.,  et~al., 2017, preprint, \href
  {http://adsabs.harvard.edu/abs/2017arXiv170904474G} {} (\mn@eprint {arXiv}
  {1709.04474})

\bibitem[\protect\citeauthoryear{{Greco} et~al.,}{{Greco}
  et~al.}{2018}]{2018ApJ...857..104G}
{Greco} J.~P.,  et~al., 2018, \mn@doi [\apj] {10.3847/1538-4357/aab842}, \href
  {https://ui.adsabs.harvard.edu/#abs/2018ApJ...857..104G} {857, 104}

\bibitem[\protect\citeauthoryear{{Greggio}, {Rejkuba}, {Gonzalez}, {Arnaboldi},
  {Iodice}, {Irwin}, {Neeser}  \& {Emerson}}{{Greggio}
  et~al.}{2014}]{2014A&A...562A..73G}
{Greggio} L.,  {Rejkuba} M.,  {Gonzalez} O.~A.,  {Arnaboldi} M.,  {Iodice} E.,
  {Irwin} M.,  {Neeser} M.~J.,   {Emerson} J.,  2014, \mn@doi [\aap]
  {10.1051/0004-6361/201322759}, \href
  {https://ui.adsabs.harvard.edu/#abs/2014A&A...562A..73G} {562, A73}

\bibitem[\protect\citeauthoryear{Haines}{Haines}{2010}]{helit}
Haines T. S.~F.,  2010, \url{https://github.com/thaines/helit}

\bibitem[\protect\citeauthoryear{Hastie}{Hastie}{1984}]{HastieThesis}
Hastie T.,  1984, PhD thesis, Stanford Linear Accelerator Center, Stanford
  University, \url {http://pca.narod.ru/HastieThesis.htm}

\bibitem[\protect\citeauthoryear{Hastie \& Stuetzle}{Hastie \&
  Stuetzle}{1989}]{Hastie1989}
Hastie T.,  Stuetzle W.,  1989, \mn@doi [Journal of the American Statistical
  Association] {10.1080/01621459.1989.10478797}, 84, 502

\bibitem[\protect\citeauthoryear{{He}, {Alam}, {Ferraro}, {Chen}  \& {Ho}}{{He}
  et~al.}{2017}]{2017arXiv170902543H}
{He} S.,  {Alam} S.,  {Ferraro} S.,  {Chen} Y.-C.,   {Ho} S.,  2017, preprint,
  \href {https://ui.adsabs.harvard.edu/#abs/2017arXiv170902543H} {p.
  arXiv:1709.02543} (\mn@eprint {arXiv} {1709.02543})

\bibitem[\protect\citeauthoryear{{Helmi} \& {White}}{{Helmi} \&
  {White}}{1999}]{1999MNRAS.307..495H}
{Helmi} A.,  {White} S. D.~M.,  1999, \mn@doi [\mnras]
  {10.1046/j.1365-8711.1999.02616.x}, \href
  {https://ui.adsabs.harvard.edu/#abs/1999MNRAS.307..495H} {307, 495}

\bibitem[\protect\citeauthoryear{{Hendel} \& {Johnston}}{{Hendel} \&
  {Johnston}}{2015}]{2015MNRAS.454.2472H}
{Hendel} D.,  {Johnston} K.~V.,  2015, \mn@doi [\mnras]
  {10.1093/mnras/stv2035}, \href
  {https://ui.adsabs.harvard.edu/#abs/2015MNRAS.454.2472H} {454, 2472}

\bibitem[\protect\citeauthoryear{{Hernquist} \& {Ostriker}}{{Hernquist} \&
  {Ostriker}}{1992}]{1992ApJ...386..375H}
{Hernquist} L.,  {Ostriker} J.~P.,  1992, \mn@doi [\apj] {10.1086/171025},
  \href {https://ui.adsabs.harvard.edu/#abs/1992ApJ...386..375H} {386, 375}

\bibitem[\protect\citeauthoryear{{Hood}, {Kannappan}, {Stark}, {Dell'Antonio},
  {Moffett}, {Eckert}, {Norris}  \& {Hendel}}{{Hood}
  et~al.}{2018}]{2018ApJ...857..144H}
{Hood} C.~E.,  {Kannappan} S.~J.,  {Stark} D.~V.,  {Dell'Antonio} I.~P.,
  {Moffett} A.~J.,  {Eckert} K.~D.,  {Norris} M.~A.,   {Hendel} D.,  2018,
  \mn@doi [\apj] {10.3847/1538-4357/aab719}, \href
  {http://adsabs.harvard.edu/abs/2018ApJ...857..144H} {857, 144}

\bibitem[\protect\citeauthoryear{Hunter}{Hunter}{2007}]{Hunter:2007}
Hunter J.~D.,  2007, \mn@doi [Computing In Science \& Engineering]
  {10.1109/MCSE.2007.55}, 9, 90

\bibitem[\protect\citeauthoryear{{Ibata}, {Gilmore}  \& {Irwin}}{{Ibata}
  et~al.}{1994}]{1994Natur.370..194I}
{Ibata} R.~A.,  {Gilmore} G.,   {Irwin} M.~J.,  1994, \mn@doi [\nat]
  {10.1038/370194a0}, \href
  {https://ui.adsabs.harvard.edu/#abs/1994Natur.370..194I} {370, 194}

\bibitem[\protect\citeauthoryear{{Ibata}, {Irwin}, {Lewis}, {Ferguson}  \&
  {Tanvir}}{{Ibata} et~al.}{2001}]{2001Natur.412...49I}
{Ibata} R.,  {Irwin} M.,  {Lewis} G.,  {Ferguson} A. M.~N.,   {Tanvir} N.,
  2001, \nat, \href {https://ui.adsabs.harvard.edu/#abs/2001Natur.412...49I}
  {412, 49}

\bibitem[\protect\citeauthoryear{{Ibata}, {Martin}, {Irwin}, {Chapman},
  {Ferguson}, {Lewis}  \& {McConnachie}}{{Ibata}
  et~al.}{2007}]{2007ApJ...671.1591I}
{Ibata} R.,  {Martin} N.~F.,  {Irwin} M.,  {Chapman} S.,  {Ferguson} A.~M.~N.,
  {Lewis} G.~F.,   {McConnachie} A.~W.,  2007, \mn@doi [\apj] {10.1086/522574},
  \href {https://ui.adsabs.harvard.edu/#abs/2007ApJ...671.1591I} {671, 1591}

\bibitem[\protect\citeauthoryear{{Ibata} et~al.,}{{Ibata}
  et~al.}{2014}]{2014ApJ...780..128I}
{Ibata} R.~A.,  et~al., 2014, \mn@doi [\apj] {10.1088/0004-637X/780/2/128},
  \href {https://ui.adsabs.harvard.edu/#abs/2014ApJ...780..128I} {780, 128}

\bibitem[\protect\citeauthoryear{{Johnston}, {Bullock}, {Sharma}, {Font},
  {Robertson}  \& {Leitner}}{{Johnston} et~al.}{2008}]{2008ApJ...689..936J}
{Johnston} K.~V.,  {Bullock} J.~S.,  {Sharma} S.,  {Font} A.,  {Robertson}
  B.~E.,   {Leitner} S.~N.,  2008, \mn@doi [\apj] {10.1086/592228}, \href
  {https://ui.adsabs.harvard.edu/#abs/2008ApJ...689..936J} {689, 936}

\bibitem[\protect\citeauthoryear{Jones, Oliphant, Peterson  et~al.}{Jones
  et~al.}{2001}]{scipy}
Jones E.,  Oliphant T.,  Peterson P.,   et~al., 2001, {SciPy}: Open source
  scientific tools for {Python}, \url {http://www.scipy.org/}

\bibitem[\protect\citeauthoryear{{Kado-Fong} et~al.,}{{Kado-Fong}
  et~al.}{2018}]{2018arXiv180505970K}
{Kado-Fong} E.,  et~al., 2018, preprint, \href
  {https://ui.adsabs.harvard.edu/#abs/2018arXiv180505970K} {p.
  arXiv:1805.05970} (\mn@eprint {arXiv} {1805.05970})

\bibitem[\protect\citeauthoryear{{Karademir}, {Remus}, {Burkert}, {Dolag},
  {Hoffmann}, {Moster}, {Steinwandel}  \& {Zhang}}{{Karademir}
  et~al.}{2018}]{2018arXiv180810454K}
{Karademir} G.~S.,  {Remus} R.-S.,  {Burkert} A.,  {Dolag} K.,  {Hoffmann}
  T.~L.,  {Moster} B.~P.,  {Steinwandel} U.,   {Zhang} J.,  2018, preprint,
  \href {https://ui.adsabs.harvard.edu/#abs/2018arXiv180810454K} {p.
  arXiv:1808.10454} (\mn@eprint {arXiv} {1808.10454})

\bibitem[\protect\citeauthoryear{{K{\"u}pper}, {Kroupa}, {Baumgardt}  \&
  {Heggie}}{{K{\"u}pper} et~al.}{2010}]{2010MNRAS.401..105K}
{K{\"u}pper} A. H.~W.,  {Kroupa} P.,  {Baumgardt} H.,   {Heggie} D.~C.,  2010,
  \mn@doi [\mnras] {10.1111/j.1365-2966.2009.15690.x}, \href
  {https://ui.adsabs.harvard.edu/#abs/2010MNRAS.401..105K} {401, 105}

\bibitem[\protect\citeauthoryear{{LSST Science Collaboration} et~al.,}{{LSST
  Science Collaboration} et~al.}{2009}]{2009arXiv0912.0201L}
{LSST Science Collaboration} et~al., 2009, preprint, \href
  {https://ui.adsabs.harvard.edu/#abs/2009arXiv0912.0201L} {p. arXiv:0912.0201}
  (\mn@eprint {arXiv} {0912.0201})

\bibitem[\protect\citeauthoryear{Lam, Pitrou  \& Seibert}{Lam
  et~al.}{2015}]{numba}
Lam S.~K.,  Pitrou A.,   Seibert S.,  2015, in Proceedings of the Second
  Workshop on the LLVM Compiler Infrastructure in HPC. LLVM '15.
ACM, New York, NY, USA, pp 7:1--7:6, \mn@doi{10.1145/2833157.2833162}, \url
  {http://doi.acm.org/10.1145/2833157.2833162}

\bibitem[\protect\citeauthoryear{{Majewski}, {Skrutskie}, {Weinberg}  \&
  {Ostheimer}}{{Majewski} et~al.}{2003}]{2003ApJ...599.1082M}
{Majewski} S.~R.,  {Skrutskie} M.~F.,  {Weinberg} M.~D.,   {Ostheimer} J.~C.,
  2003, \mn@doi [\apj] {10.1086/379504}, \href
  {https://ui.adsabs.harvard.edu/#abs/2003ApJ...599.1082M} {599, 1082}

\bibitem[\protect\citeauthoryear{{Mart{\'\i}nez-Delgado}
  et~al.,}{{Mart{\'\i}nez-Delgado} et~al.}{2010}]{2010AJ....140..962M}
{Mart{\'\i}nez-Delgado} D.,  et~al., 2010, \mn@doi [\aj]
  {10.1088/0004-6256/140/4/962}, \href
  {https://ui.adsabs.harvard.edu/#abs/2010AJ....140..962M} {140, 962}

\bibitem[\protect\citeauthoryear{{McConnachie} et~al.,}{{McConnachie}
  et~al.}{2009}]{2009Natur.461...66M}
{McConnachie} A.~W.,  et~al., 2009, \mn@doi [\nat] {10.1038/nature08327}, \href
  {https://ui.adsabs.harvard.edu/#abs/2009Natur.461...66M} {461, 66}

\bibitem[\protect\citeauthoryear{{McConnachie} et~al.,}{{McConnachie}
  et~al.}{2018}]{2018arXiv181008234M}
{McConnachie} A.~W.,  et~al., 2018, preprint, \href
  {https://ui.adsabs.harvard.edu/#abs/2018arXiv181008234M} {p.
  arXiv:1810.08234} (\mn@eprint {arXiv} {1810.08234})

\bibitem[\protect\citeauthoryear{{Mihos}, {Durrell}, {Feldmeier}, {Harding}  \&
  {Watkins}}{{Mihos} et~al.}{2018}]{2018arXiv180606828M}
{Mihos} J.~C.,  {Durrell} P.~R.,  {Feldmeier} J.~J.,  {Harding} P.,   {Watkins}
  A.~E.,  2018, preprint, \href
  {https://ui.adsabs.harvard.edu/#abs/2018arXiv180606828M} {p.
  arXiv:1806.06828} (\mn@eprint {arXiv} {1806.06828})

\bibitem[\protect\citeauthoryear{{Miskolczi}, {Bomans}  \&
  {Dettmar}}{{Miskolczi} et~al.}{2011}]{2011A&A...536A..66M}
{Miskolczi} A.,  {Bomans} D.~J.,   {Dettmar} R.~J.,  2011, \mn@doi [\aap]
  {10.1051/0004-6361/201116716}, \href
  {https://ui.adsabs.harvard.edu/#abs/2011A&A...536A..66M} {536, A66}

\bibitem[\protect\citeauthoryear{{Monachesi}, {Bell}, {Radburn-Smith},
  {Bailin}, {de Jong}, {Holwerda}, {Streich}  \& {Silverstein}}{{Monachesi}
  et~al.}{2016}]{2016MNRAS.457.1419M}
{Monachesi} A.,  {Bell} E.~F.,  {Radburn-Smith} D.~J.,  {Bailin} J.,  {de Jong}
  R.~S.,  {Holwerda} B.,  {Streich} D.,   {Silverstein} G.,  2016, \mn@doi
  [\mnras] {10.1093/mnras/stv2987}, \href
  {https://ui.adsabs.harvard.edu/#abs/2016MNRAS.457.1419M} {457, 1419}

\bibitem[\protect\citeauthoryear{{Morales}, {Mart{\'\i}nez-Delgado}, {Grebel},
  {Cooper}, {Javanmardi}  \& {Miskolczi}}{{Morales}
  et~al.}{2018}]{2018arXiv180403330M}
{Morales} G.,  {Mart{\'\i}nez-Delgado} D.,  {Grebel} E.~K.,  {Cooper} A.~P.,
  {Javanmardi} B.,   {Miskolczi} A.,  2018, preprint, \href
  {https://ui.adsabs.harvard.edu/#abs/2018arXiv180403330M} {p.
  arXiv:1804.03330} (\mn@eprint {arXiv} {1804.03330})

\bibitem[\protect\citeauthoryear{{Mouhcine}, {Ibata}  \& {Rejkuba}}{{Mouhcine}
  et~al.}{2010}]{2010ApJ...714L..12M}
{Mouhcine} M.,  {Ibata} R.,   {Rejkuba} M.,  2010, \mn@doi [\apj]
  {10.1088/2041-8205/714/1/L12}, \href
  {https://ui.adsabs.harvard.edu/#abs/2010ApJ...714L..12M} {714, L12}

\bibitem[\protect\citeauthoryear{{Navarro}, {Frenk}  \& {White}}{{Navarro}
  et~al.}{1997}]{1997ApJ...490..493N}
{Navarro} J.~F.,  {Frenk} C.~S.,   {White} S. D.~M.,  1997, \mn@doi [\apj]
  {10.1086/304888}, \href
  {https://ui.adsabs.harvard.edu/#abs/1997ApJ...490..493N} {490, 493}

\bibitem[\protect\citeauthoryear{{Newberg} et~al.,}{{Newberg}
  et~al.}{2002}]{2002ApJ...569..245N}
{Newberg} H.~J.,  et~al., 2002, \mn@doi [\apj] {10.1086/338983}, \href
  {https://ui.adsabs.harvard.edu/#abs/2002ApJ...569..245N} {569, 245}

\bibitem[\protect\citeauthoryear{{Okamoto}, {Arimoto}, {Ferguson}, {Bernard},
  {Irwin}, {Yamada}  \& {Utsumi}}{{Okamoto} et~al.}{2015}]{2015ApJ...809L...1O}
{Okamoto} S.,  {Arimoto} N.,  {Ferguson} A. M.~N.,  {Bernard} E.~J.,  {Irwin}
  M.~J.,  {Yamada} Y.,   {Utsumi} Y.,  2015, \mn@doi [\apj]
  {10.1088/2041-8205/809/1/L1}, \href
  {https://ui.adsabs.harvard.edu/#abs/2015ApJ...809L...1O} {809, L1}

\bibitem[\protect\citeauthoryear{Ozertem \& Erdogmus}{Ozertem \&
  Erdogmus}{2011}]{SCMS}
Ozertem U.,  Erdogmus D.,  2011, J. Mach. Learn. Res., 12, 1249

\bibitem[\protect\citeauthoryear{Patra \& Sen}{Patra \& Sen}{2016}]{patra2016}
Patra R.~K.,  Sen B.,  2016, \mn@doi [Journal of the Royal Statistical Society:
  Series B (Statistical Methodology)] {10.1111/rssb.12148}, 78, 869

\bibitem[\protect\citeauthoryear{{Pe{\~n}arrubia}, {Navarro}  \&
  {McConnachie}}{{Pe{\~n}arrubia} et~al.}{2008}]{2008ApJ...673..226P}
{Pe{\~n}arrubia} J.,  {Navarro} J.~F.,   {McConnachie} A.~W.,  2008, \mn@doi
  [\apj] {10.1086/523686}, \href
  {https://ui.adsabs.harvard.edu/#abs/2008ApJ...673..226P} {673, 226}

\bibitem[\protect\citeauthoryear{P\'erez \& Granger}{P\'erez \&
  Granger}{2007}]{ipython}
P\'erez F.,  Granger B.~E.,  2007, \mn@doi [Computing in Science and
  Engineering] {10.1109/MCSE.2007.53}, 9, 21

\bibitem[\protect\citeauthoryear{{Pop}, {Pillepich}, {Amorisco}  \&
  {Hernquist}}{{Pop} et~al.}{2017}]{2017arXiv170606102P}
{Pop} A.-R.,  {Pillepich} A.,  {Amorisco} N.~C.,   {Hernquist} L.,  2017,
  preprint, \href {https://ui.adsabs.harvard.edu/#abs/2017arXiv170606102P} {p.
  arXiv:1706.06102} (\mn@eprint {arXiv} {1706.06102})

\bibitem[\protect\citeauthoryear{{Quinn}}{{Quinn}}{1984}]{1984ApJ...279..596Q}
{Quinn} P.~J.,  1984, \mn@doi [\apj] {10.1086/161924}, \href
  {https://ui.adsabs.harvard.edu/#abs/1984ApJ...279..596Q} {279, 596}

\bibitem[\protect\citeauthoryear{{Radburn-Smith} et~al.,}{{Radburn-Smith}
  et~al.}{2011}]{2011ApJS..195...18R}
{Radburn-Smith} D.~J.,  et~al., 2011, \mn@doi [The Astrophysical Journal
  Supplement Series] {10.1088/0067-0049/195/2/18}, \href
  {https://ui.adsabs.harvard.edu/#abs/2011ApJS..195...18R} {195, 18}

\bibitem[\protect\citeauthoryear{{Rejkuba}, {Harris}, {Greggio}, {Harris},
  {Jerjen}  \& {Gonzalez}}{{Rejkuba} et~al.}{2014}]{2014ApJ...791L...2R}
{Rejkuba} M.,  {Harris} W.~E.,  {Greggio} L.,  {Harris} G.~L.~H.,  {Jerjen} H.,
    {Gonzalez} O.~A.,  2014, \mn@doi [\apj] {10.1088/2041-8205/791/1/L2}, \href
  {https://ui.adsabs.harvard.edu/#abs/2014ApJ...791L...2R} {791, L2}

\bibitem[\protect\citeauthoryear{{Sanderson} \& {Helmi}}{{Sanderson} \&
  {Helmi}}{2013}]{2013MNRAS.435..378S}
{Sanderson} R.~E.,  {Helmi} A.,  2013, \mn@doi [\mnras]
  {10.1093/mnras/stt1307}, \href
  {https://ui.adsabs.harvard.edu/#abs/2013MNRAS.435..378S} {435, 378}

\bibitem[\protect\citeauthoryear{{Smith}, {Choi}, {Lee}, {Rhee},
  {Sanchez-Janssen}  \& {Yi}}{{Smith} et~al.}{2016}]{2016ApJ...833..109S}
{Smith} R.,  {Choi} H.,  {Lee} J.,  {Rhee} J.,  {Sanchez-Janssen} R.,   {Yi}
  S.~K.,  2016, \mn@doi [\apj] {10.3847/1538-4357/833/1/109}, \href
  {https://ui.adsabs.harvard.edu/#abs/2016ApJ...833..109S} {833, 109}

\bibitem[\protect\citeauthoryear{Sobel}{Sobel}{2014}]{sobel}
Sobel I.,  2014, Presentation at Stanford A.I. Project 1968

\bibitem[\protect\citeauthoryear{{The Astropy Collaboration} et~al.,}{{The
  Astropy Collaboration} et~al.}{2018}]{2018arXiv180102634T}
{The Astropy Collaboration} et~al., 2018, preprint, \href
  {https://ui.adsabs.harvard.edu/#abs/2018arXiv180102634T} {p.
  arXiv:1801.02634} (\mn@eprint {arXiv} {1801.02634})

\bibitem[\protect\citeauthoryear{{Villalobos}, {De Lucia}, {Borgani}  \&
  {Murante}}{{Villalobos} et~al.}{2012}]{2012MNRAS.424.2401V}
{Villalobos} {\'A}.,  {De Lucia} G.,  {Borgani} S.,   {Murante} G.,  2012,
  \mn@doi [\mnras] {10.1111/j.1365-2966.2012.20667.x}, \href
  {https://ui.adsabs.harvard.edu/#abs/2012MNRAS.424.2401V} {424, 2401}

\bibitem[\protect\citeauthoryear{{Weinberg}}{{Weinberg}}{1996}]{1996ApJ...470..715W}
{Weinberg} M.~D.,  1996, \mn@doi [\apj] {10.1086/177902}, \href
  {https://ui.adsabs.harvard.edu/#abs/1996ApJ...470..715W} {470, 715}

\bibitem[\protect\citeauthoryear{{White} \& {Rees}}{{White} \&
  {Rees}}{1978}]{1978MNRAS.183..341W}
{White} S.~D.~M.,  {Rees} M.~J.,  1978, \mn@doi [\mnras]
  {10.1093/mnras/183.3.341}, \href
  {https://ui.adsabs.harvard.edu/#abs/1978MNRAS.183..341W} {183, 341}

\bibitem[\protect\citeauthoryear{Wilcoxon}{Wilcoxon}{1945}]{wilcoxon}
Wilcoxon F.,  1945, \mn@doi [Biometrics Bulletin] {10.2307/3001968}, 1, 80

\makeatother
\end{thebibliography}




\appendix
\section{More examples of SCUDS}
\label{appendix}

\begin{figure*}
	\includegraphics[width=.95\linewidth]{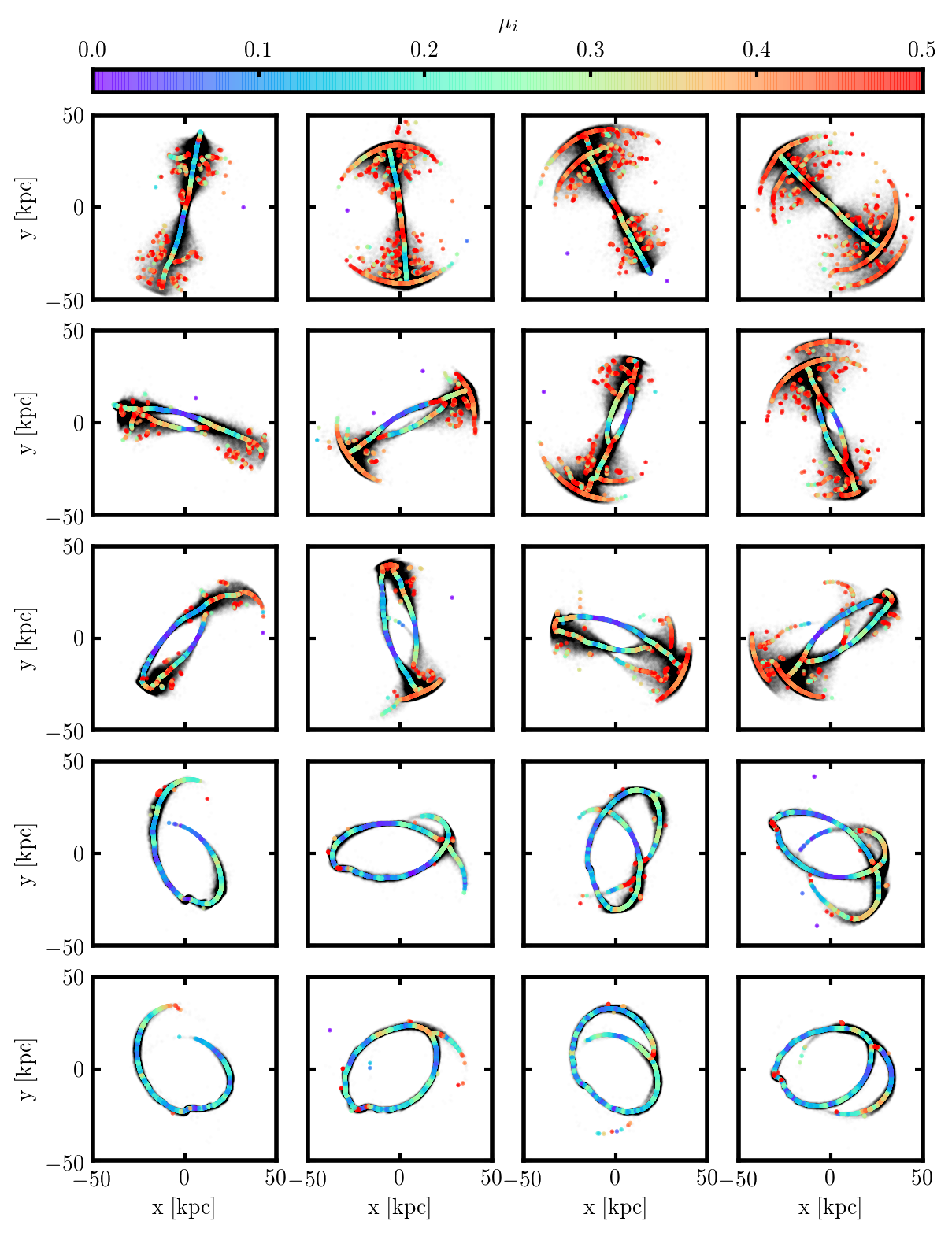}
        \caption{Examples of SCUDS for simulation snapshots with a satellite mass of  $6.5 \times 10^7\ {\rm{M_\odot}}$; see text of Appendix~\ref{appendix} for details.}
    \label{fig:1e7}
\end{figure*}

\begin{figure*}
	\includegraphics[width=.95\linewidth]{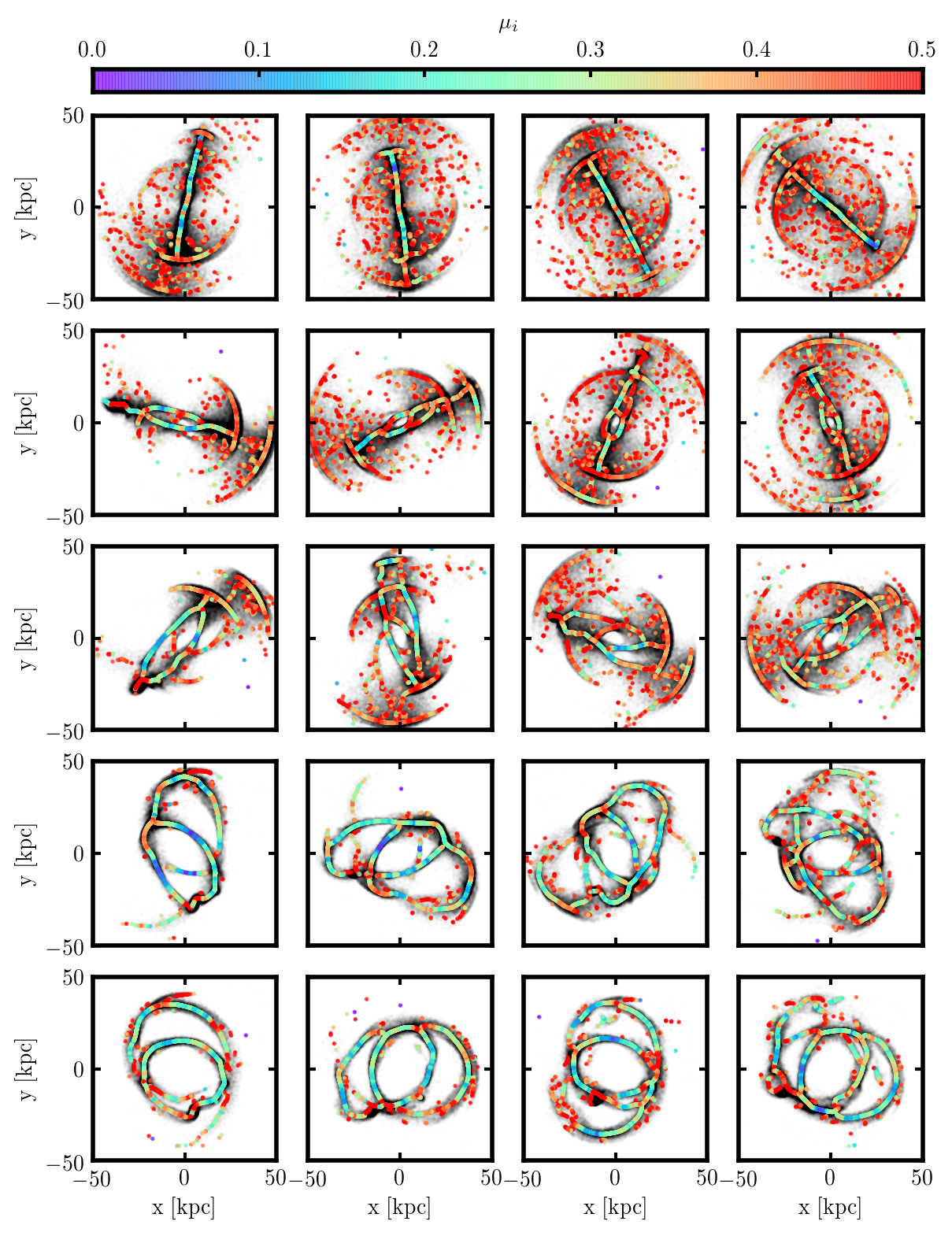}
        \caption{Examples of SCUDS for simulation snapshots with a satellite mass of  $6.5 \times 10^8\ {\rm{M_\odot}}$; see text of Appendix~\ref{appendix} for details.}
    \label{fig:1e8}
\end{figure*}

In this Appendix's Figures \ref{fig:1e7} and \ref{fig:1e8} we show additional examples of the application of SCUDS to the N--body simulations described in the text for a satellite mass of $6.5 \times 10^7\ {\rm{M_\odot}}$ and $6.5 \times 10^8\ {\rm{M_\odot}}$, respectively. In both figures the orbit of the satellite has the same energy as that of a circular orbit at 25 kpc. The four columns represent simulation snapshots after integration for 3.2, 4.0, 4.8, and 5.6 Gyr and the rows represent simulations with angular momentum equal to 0.1, 0.3, 0.5, 0.8, and 0.95 times that of the same circular orbit. As in the main text, the black points indicate all N--body particles while the larger colored points are the ridgepoints shaded by the local morphology $\mu_i$.

%


\bsp	
\label{lastpage}
\end{document}